\documentclass[preprint,showpacs,aps,prb,floatfix,endfloats]{revtex4}

\usepackage{graphicx}

\begin{document}


\title{ 
 Energy and structure of dilute hard-- and soft--sphere gases
}
\author{F.Mazzanti$^1$, A.Polls$^2$, A. Fabrocini$^3$}
\affiliation{ $^1$ Departament d'Electr\`onica, Enginyeria i Arquitectura
  La Salle, \\ 
  Pg. Bonanova 8, Universitat Ramon Llull, \\
  E-08022 Barcelona, Spain}
\affiliation{ $^2$ Departament d'Estructura i Constituents de la
  Mat\`eria, \\
  Diagonal 645, Universitat de Barcelona, \\
  E-08028 Barcelona, Spain}
\affiliation{$^3$Dipartimento di Fisica "E.Fermi", Universit\`a di Pisa, 
and INFN, Sezione di Pisa, \\ 
Via Buonarroti,2 \\
I-56100 Pisa, Italy}


\begin{abstract}

The energy and structure of dilute hard- and soft-sphere Bose gases are  systematically studied
in the framework of several many-body approaches, as the variational correlated theory,
the Bogoliubov model and the uniform limit approximation, valid in the weak interaction
regime. When possible, the results are compared with the exact diffusion Monte Carlo ones.
A Jastrow type correlation provides a good description of the systems, both hard- and soft-spheres,
if the hypernetted chain energy functional is freely minimized and the resulting Euler equation
is solved. The study of the soft-spheres potentials confirms the appearance of a dependence
of the energy on the shape of the potential at gas paremeter values of $x \sim 0.001$. For quantities other than the energy, such as the radial distribution 
functions and the momentum distributions,  
the dependence appears at any  value of $x$. The occurrence 
of a maximum in the radial
distribution function, in the momentum distribution and in the 
excitation spectrum is a natural effect of the correlations when $x$ 
increases. The asymptotic behaviors of the functions characterizing the 
structure of the systems are also investigated. 
The uniform limit approach results very easy to implement and 
provides a good description of the  soft-sphere gas. Its 
reliability improves when the interaction weakens.

\end{abstract}
\maketitle

\section {Introduction}

The study of dilute systems has met a surge of renewed interest in the
last years, following the experimental achievement of Bose-Einstein
condensation  in low density atomic gases confined in harmonic
traps.  Systems are termed as dilute when the average interparticle
distance, $r_{av}$, is much larger than the range of the interaction,
$r_0$. The main parameter characterizing the interaction in the
dilute regime is the $s$--wave scattering length, $a$, and the diluteness
condition can be expressed in terms of the density, $\rho$, as $x<<1$,
where $x=\rho a^3$ is the gas parameter.

Positive values of the scattering length correspond to an 
essentially repulsive  interaction
 at short distances.  In fact, for an infinite
repulsive barrier, with no attractive part, $a$ is just given by the
radius of the barrier itself. The hard sphere (HS) potential is
largely adopted to study low density (LD) gases with positive
scattering lengths  because of its formal simplicity. However, other
potentials, providing the same value of $a$, are indistinguishable
from the HS one at very low $x$ ({\em universality} property). 
Additional details of the interaction, giving raise to 
{\em nonuniversal} effects, 
become relevant when the density (and the gas parameter) increases,
allowing for discriminating among different potential shapes. Typical
ranges of $x$ where these differences begin to show up have been 
found\cite{boro99,sarsa02} to be $x\geq 10^{-3}$.

In the density ranges attained in Bose-Einstein condensation  experiments the local gas
parameter may well exceed this value, exploiting the large variation of
the scattering length in the vicinity of a Fehsbach
resonance\cite{cornish00}. In order to quantitatively study 
this regime it is compulsory to check the reliability of the 
theories adopted in the analysis. 
A first, and much needed step in this direction, consists in 
understanding the properties of the underlying homogeneous gas 
of bosons at T=0 temperature.  
This is a time honored subject of many--body physics,
addressed by a variety of methods, such as perturbative expansions,
Monte Carlo (MC) type samplings, variational theories, and so on.

Expansions in $x$ may be suited to study dilute  
systems whose interaction can be safely described 
in terms of the $s$--wave scattering length. They can be 
derived in the framework of standard perturbation theories, 
built on the basis constituted by the
ground state of a non interacting Bose fluid and its excited states,
obtained by promoting particles from the zero momentum condensed state
to the non--zero ones.  Infinite sums of ladder diagrams must be
accomplished for strongly repulsive potentials.  This procedure results
 in the well known Lee and Yang  low density  expansion for the 
energy per particle of a homogeneous gas of bosons\cite{lee57}, 
\begin{equation}
{E\over N} = \left( {\hbar^2 \over {2 m a^2}} \right)\, 
4\pi x \, \left[ 1 + {128 \over 15} \sqrt{{x\over \pi}} \right] \, .
\label{E_LY}
\end{equation}
Low order perturbative diagrams can be retained in the weakly interacting 
Bose gas\cite{fetter} characterized by
an interaction having a finite Fourier transform. 

Alternatively, the interaction may be mimicked in the low density region 
by a delta--shaped pseudopotential\cite{huang}, related to $a$ by 
\begin{equation}
V(r)=4\pi a \hbar^2 \delta ({\bf r})/m \ , 
\label{delta}
\end{equation}
which reproduces the value of the scattering length 
in the Born approximation:
\begin{equation}
a_B ={m \over {4\pi \hbar^2}} \int d{\bf r} \, V(r) \, .
\label{Born}
\end{equation}
This approach is followed in the Bogoliubov theory\cite{bogo}, 
by introducing a model hamiltonian containing the pseudopotential. 
A canonical transformation of the BCS type allows for a partial 
diagonalization of the model hamiltonian, leading to a  
 $x$--expansion as in (\ref{E_LY}) for the ground--state 
energy.

Nonuniversal effects have been studied in effective field theory in 
Ref.\cite{braaten01}, and found to depend, at the leading orders, on the 
effective range, $r_s$, and on a three--body contact force parameter, $c$, 
determined in this reference by comparison with the exact diffusion Monte 
Carlo results of Ref.\cite{boro99}.



The bosonic many--body Schroedinger equation may be solved, for any 
potential, by Monte Carlo based methods, as the Green's Function 
Monte Carlo 
 and the Diffusion Monte Carlo 
(DMC)
 ones. Both approaches are exact, apart from
statistical errors, and provide essential benchmarks to test 
the reliability of other theories, at least in those cases where 
the numerical  accuracy allows for an unbiased analysis of 
the results.

The variational approach is carried on within the Correlated Basis 
Functions (CBF) perturbation theory, and represents its zeroth 
order. The correlated ground state wave function for $N$ interacting 
particles is obtained by acting 
with a many--body correlation operator on the non--interacting 
ground state wave function\cite{feenberg}:
\begin{equation}
\Psi_0(1,2,\ldots , N)=
{\cal F}(1,2,\ldots , N)
\Phi_0(1,2,\ldots , N) \, .
\label{CBF}
\end{equation}
The operator ${\cal F}(1,2,\ldots , N)$ is meant to take into account
the spatial correlations induced by the interaction on the free wave
function $\Phi_0(1,2,\ldots , N)$ (for homogeneous bosons $\Phi_0=1$).
For instance, in the HS case the correlation operator prevents the
distance between any pair of particles from being smaller than the
core radius, so that the wave function vanishes for these
configurations.  The theory is variational in the sense that ${\cal
F}$ is determined by minimizing the ground state energy. A correlated
perturbation theory may be constructed by applying the correlation
operator to the non--interacting excited states.

The weakly interacting Bose fluid has been recently studied within 
the variational method by means of the Independent Pair Correlations 
(IPC) approach\cite{sarsa02}. The IPC wave function is written as 
\begin{equation}
\Psi_{IPC}(1,2,\ldots , N)=
1+\sum_{i<j} h(r_{ij})
+{1 \over 2} \sum_{(i<j)\neq (l<m)}h(r_{ij})h(r_{lm})+\ldots \, ,
\label{IPC}
\end{equation}
where the two--body correlation functions, $h(r)$, always act on
non--overlapping, independent pairs. The structure of a Bose system
described by the IPC wave function has been analyzed by
means of an expansion in cluster diagrams.  Renormalized HyperNetted
Chain (HNC) integral equations exactly sum all the diagrams and have
been used to compute the energy, distribution function, momentum
distribution and pairing function of the bulk boson system of soft
spheres. The Bogoliubov transformation itself is known to generate a
wave function having the IPC structure\cite{huang}, and a careful
comparison between the variational and Bogoliubov approaches has been
performed in Ref.\cite{sarsa02}, as well as with the outcomes of the
Diffusion Monte Carlo  and low density  theories, for soft sphere (SS) and gaussian like
potentials.

The IPC wave function fails for a hard--sphere fluid, since all 
pairs need to be simultaneously correlated. This task may be achieved 
by a Jastrow correlated wave function\cite{feenberg} 
\begin{equation}
\Psi_{J}(1,2,\ldots , N)=
\prod_{i<j} [ 1+h(r_{ij}) ] \, ,
\label{Jastrow}
\end{equation}
where the productory runs over $all$ pairs. The Jastrow ansatz is
clearly richer than the IPC one, however it has the disvantage that
the HNC equations can be only approximately solved, since a class of
diagrams (the $elementary$ diagrams) is not summable in a closed
way\cite{HNC}.  The approximations in the solution of the Jastrow--HNC
equations are not expected to be quantitatively relevant in the low
density regime, however the computed energy is, in principle, no
longer a rigorous upperbound to the true one, in contrast with the IPC
case.

In this paper  we consider a system of $N$ spinless 
bosons of mass $m$ in a volume $\Omega$, described by the hamiltonian
\begin{equation}
H = -{\hbar^2 \over 2m} \sum_{j=1}^N \nabla^2_j + 
\sum_{1=i<j}^N V(r_{ij})  \,
\label{hamil}
\end{equation}
where $V(r)$ is a two-body, spherically symmetric potential, 
in the thermodynamic limit ( $N$ and $\Omega$ $ \rightarrow \infty$,  
keeping the density, $\rho = N/\Omega$, constant).

Two different representative choices for 
the potential are studied: the hard-sphere  potential,
\begin{equation}
V(r) = \left\{
\begin{array}{ll}
  \infty  & \,\,\,\,\, r<a \\
  0  & \,\,\,\,\, r> a \ ,
\end{array}
\right. 
\label{vpoths}
\end{equation}
where the diameter $a$ of the hard sphere 
coincides with the $s$--wave scattering length, and a soft sphere
 potential,
\begin{equation}
V(r) = \left\{
\begin{array}{ll}
  V_0 > 0 & \,\,\,\,\, r<R \\
  0  & \,\,\,\,\, r> R \ ,
\end{array}
\right. 
\label{vpotss}
\end{equation}
whose $s$-wave scattering length is given by
\begin{equation}
a=R \left [1-\tanh(K_0 R)/(K_0 R)\right ],
\label{length}
\end{equation}
with  
$K_0^2 = V_0 m/\hbar^2$. 

The $optimal$ Jastrow correlated wave function, obtained by the
minimization of the energy per particle through the solution of the
HNC Euler equation \cite{kro98}, is used along this work. 
 We compute the radial distribution
function, $g(r)$, the static structure function, $S(k)$, and the
momentum distribution, $n(k)$, for HS and SS models with identical
scattering lengths, to ascertain the dependence on the potential form
along the gas parameter.  We work in the HNC/0 approximation, which
amounts to disregard the contribution from the elementary diagrams and
whose accuracy is tested by comparison with the exact DMC results.
Moreover, the Jastrow theory is compared with the LD expansion and
with the Bogoliubov and IPC theories for the SS case.  In addition,
the SS potential is examined in the so called {\it uniform limit}
(UL)\cite{feenberg}, defined by assuming $\vert g(r)-1|<< 1$ for all $r$. 
Special emphasis is devoted to the analysis of several asymptotic
behaviors. This is an interesting issue that cannot be 
fully addressed with DMC methods, since the limited size 
of the simulation box strongly limits the possibility of 
studying effects related to long range structures. 
 
The plan of the paper is as follows: the HNC/0 and Euler equations 
are shortly revisited in Section II; the uniform limit for the 
SS potential and its connection with the Random Phase Approximation (RPA) 
theory are discussed 
in Section III; Section IV present the results for the HS and 
SS models and summary and conclusions are given in Section V. 

\section{HNC theory}

The exact wave function of a homogenous, interacting Bose
system can be written as the product of up to $N$--body correlation
factors\cite{feenberg}
\begin{equation}
\Psi_0(1,2, \ldots,N) = 
\prod_{i<j}^N f_2(r_{ij}) \prod_{k<l<m}^N 
f_3({\bf r}_{kl},{\bf r}_{km},{\bf r}_{lm})
\cdots \ ,
\label{Feenberg_wf}
\end{equation}
where $r_{ij}=\mid{\bf r}_i-{\bf r}_j\mid$ and $f_n>0$ for all $n$.
High density and strongly interacting systems, 
like atomic liquid $^4$He,
are accurately described by keeping only two-- and three--body 
correlations ($f_{n>3}=1$). Moreover, a proper choice of $f_2$ already 
contributes by $\sim 70\%$ to the $^4$He energy\cite{HNC}.
So, for weak interactions 
and/or low density fluids, the simpler Jastrow correlated wave function 
of Eq.(\ref{Jastrow}) may largely be enough to capture the essential 
features of the exact ground state and to provide a quantitatively 
correct description.

\subsection{Radial distribution function and Euler equation}

The {\it optimal} Jastrow correlation function is obtained by 
minimizing, without restrictions, the expectation value 
of the hamiltonian (\ref{hamil}),
 giving the ground state energy, 
\begin{equation}
E[f_2]=
{\left\langle \Psi_0 \mid H \mid \Psi_0 \right\rangle \over
\left\langle \Psi_0 \mid \Psi_0 \right\rangle}  \, .
\label{Energy}
\end{equation}
This is accomplished  by 
solving the Euler--Lagrange equation \cite{camp}
\begin{equation}
{\delta E[f_2] \over \delta f_2(r)} = 0 \, .
\label{Euler}
\end{equation}
The link between the energy and $f_2$ is provided by the radial 
distribution function, $g(r)$,
\begin{equation}
g(r_{12})=
{{N(N-1)}\over \rho^2}
{
{\int d{\bf r}_3 d{\bf r}_4 \ldots d{\bf r}_N 
\vert \Psi_0 \vert ^2}
\over
{\int d{\bf r}_1 d{\bf r}_2 \ldots d{\bf r}_N 
\vert \Psi_0 \vert ^2}
}\, ,
\label{RDF}
\end{equation}
as, in  fact,
\begin{equation}
{E\over N} = \frac {1}{2} \rho \int d{\bf r}_{12} \,g(r_{12}) 
\,\left [V(r_{12}) -
\frac{\hbar^2}{2m} \nabla^2 \ln f_2(r_{12}) \right ] \ .
\label{energy}
\end{equation}
The Jackson--Feenberg identity\cite{feenberg} has been used to derive 
Eq.(\ref{energy}). In turn, the radial distribution function can be computed by solving 
the HyperNetted Chain equations:
\begin{eqnarray}
g(r_{12}) & = & f_2^2(r_{12}) \, e^{N(r_{12})+E(r_{12})}
\nonumber \\ 
N(r_{12}) & = & \rho \int d {\bf r}_3 
[g(r_{13}) -1 ]
[g(r_{32}) -1 - N(r_{32})] \, ,
\label{HNC}
\end{eqnarray}
where $N(r)$ and $E(r)$ are the sum of the nodal and
elementary diagrams, respectively\cite{HNC}. The function $E(r)$ is 
an input to the theory, and the solution of the HNC equations depends 
on its choice. In the HNC/0 scheme this function is set to be 
zero. This seemingly drastic approximation is, however, reliable 
at low densities since the diagrams contributing to $E(r)$, being 
highly interconnected, are relevant mostly in the large density regions. 

By inverting the relations (\ref{HNC}) it is possible to express the 
energy as a functional of the radial distribution function, and formally rewrite the Euler 
equation as:
\begin{equation}
{\delta E[g] \over \delta g(r)} = 0 \, .
\label{Euler_g}
\end{equation}
If this procedure is carried on in $r$--space, the outcome is 
an integro--differential equation for $g(r)$\cite{HNC}. 
Alternatively, the energy can be written in terms of the static 
structure function, $S(k)$, 
defined as the Fourier transform of the radial distribution function, 
\begin{equation}
S(k) = 1 + \rho \int d{\bf r} \,e^{i {\bf k}\cdot {\bf r}} (g(r) -1)
\ .
\label{Structure}
\end{equation}
Variation of $E[S]$ leads to the equations\cite{kro98}:
\begin{equation}
S(k) = { t(k) \over \sqrt{ t^2(k) + 2 V_{ph}(k) t(k) } } \ ,
\label{S_Euler}
\end{equation}
with $t(k)=\hbar^2 k^2/2m$ and $V_{ph}(k)$ the so called {\em
particle--hole interaction}. In $r$--space and in HNC/0, 
\begin{equation}
V_{ph}(r) = g(r) V(r) + 
{\hbar^2 \over m} \left| \nabla\sqrt{g(r)} \right|^2 + 
\left[ g(r) - 1 \right] \omega_I(r) \ ,
\label{Vph}
\end{equation}
where the $k$--space induced interaction, $\omega_I(k)$, is 
\begin{equation}
\omega_I(k) = -{1\over 2} t(k) 
{ \left( 2 S(k)+1 \right) \left( S(k) - 1 \right)^2 \over S^2(k) } \ .
\label{Induced}
\end{equation}
Eqs.(\ref{S_Euler}), (\ref{Vph}) and 
(\ref{Induced}) are a set of nonlinear coupled equations to be 
solved iteratively. 
Finally, the knowledge of $S(k)$ (or $g(r)$) allows to find 
the optimal Jastrow correlation function by inversion of the 
HNC/0 equations.

In the HS model the Euler formalism simplifies, since 
$V_{ph}(r<a)=-\omega_I(r)$ and 
 $V_{ph}(r>a)= \hbar^2/m \left| \nabla\sqrt{g(r)} \right|^2 + 
[ g(r) - 1 ] \omega_I(r)$.

\subsection{Momentum distribution}

The one--body density matrix  for an homogeneous Bose fluid,
\begin{equation}
\rho_1({\bf r}_1,{\bf r}_{1'}) =  
\rho_1(r_{11'}) = N 
{ \int d{\bf r}_2 d{\bf r}_3\cdots d{\bf r}_N 
\Psi_0(1,2, \ldots, N) 
\Psi_0(1',2, \ldots, N) 
 \over 
\int d{\bf r}_1 \cdots d{\bf r}_N 
\vert \Psi_0 \vert ^2 } \ ,
\label{OBDM}
\end{equation}
contains essential information about the depletion of the 
condensate  in interacting systems and the consequent 
finite occupation of single particle states carrying non--zero 
momentum. Its diagonal part coincides with the one--body 
density, and in an homogeneous system $\rho_1(r=0)=\rho$. 
The condensate fraction, $n_0$, ($i.e.$ the fractional 
occupation of the $k=0$ momentum state) is related to the 
long range order of the one--body density matrix by:  
$n_0=\rho_1(r \rightarrow \infty)/\rho$. 

The associated momentum distribution, $n(k)$, is obtained 
through the Fourier transform of the one--body density matrix,
\begin{equation}
n(k) = (2\pi)^3 \rho n_0 \delta({\bf k})  + \int d{\bf r} 
\, \exp{(i {\bf k}\cdot {\bf r})} 
\, [\rho_1(r) -\rho n_0] \, .
\label{MD}
\end{equation}
The momentum distribution is normalized as
\begin{equation}
1 = {1\over (2\pi)^3\rho} \int d{\bf k}\, n(k) \ ,
\label{norm}
\end{equation}
while  the kinetic energy per particle can be obtained by $n(k)$ through 
\begin{equation}
{T \over N} = {1\over (2\pi)^3\rho} \int d{\bf k}\,
{\hbar^2 k^2 \over 2m}\, n(k) \ .
\label{kinetic}
\end{equation}

The HNC/0 scheme has been extended to evaluate the one--body density matrix for a 
Jastrow correlated wave function\cite{fantoni78}. 
As a consequence, one gets
\begin{equation}
{{\rho_1(r)} \over \rho} =  n_0 \, e^{N_{ww}(r)} \ ,
\label{OBDM_HNC}
\end{equation}
where the new nodal function, $N_{ww}(r)$, is given by 
\begin{equation}
N_{ww}(r_{12}) = \rho \int d{\bf r}_3  
[g_{wd}(r_{13}) -1 ]
[g_{dw}(r_{32}) -1 - N_{dw}(r_{32})] \, .
\label{Nww}
\end{equation}
In the HNC/0 scheme, the functions $g_{wd}$ and $N_{wd}$ are solutions of the set of
coupled equations
\begin{eqnarray}
g_{wd}(r) & = & f(r) \, e^{N_{wd}(r)} 
\nonumber \\
N_{wd}(r_{12}) & = & \rho \int d{\bf r}_3  
[g_{wd}(r_{13}) -1 ]
[g(r_{32}) -1 - N(r_{32})] \, ,
\label{Nwd}
\end{eqnarray}
with $g_{wd}(r_{12})=g_{dw}(r_{21})$ and $N_{wd}(r_{12})=N_{dw}(r_{21})$.

The condensate fraction, $n_0$, is given in terms 
of the vertex factors, $R_w$ and $R_d$, as
\begin{equation}
n_0 = e^{2 R_w -  R_d} \ ,
\label{n0}
\end{equation}
where
\begin{equation}
R_w = \rho \int d{\bf r} \, [g_{wd}(r)-1-N_{wd}(r)]
- {\rho \over 2}\int d{\bf r} \,
 [g_{wd}(r)-1] \, N_{wd}(r) \, ,
\label{Rw}
\end{equation}
and $R_d$ is obtained by substituting in (\ref{Rw}) 
 $g_{wd}\rightarrow g$ and $N_{wd}\rightarrow N$.

\section{Soft spheres in the uniform limit approximation}

The formalism presented in the previous section is independent on
the shape of the potential. Hence it is equally well suited to 
analyze both HS and SS systems. 
However, everywhere bounded potentials, as the SS one, allow
for alternative calculations complementing the HNC results. 
A  simple  estimate of the energy, in this case, 
is provided by first order perturbation theory, which yields an 
upper--bound to the exact ground state energy:
\begin{equation}
\frac {E_1(\rho)} {N} =  \frac {\langle \Phi_0 \mid H\mid \Phi_0
\rangle }{N} =\frac{1}{2} \rho V_0 \frac {4}{3} \pi R^3 = \frac{1}{2}
\tilde V(0) \ ,
\label{ener-upper}
\end{equation}
where $\tilde V$ is the Fourier transform of the potential and
$\phi_0 = 1/ \Omega^{N/2}$ is the wave function of the
corresponding free system, with all particles occupying the zero momentum
state. The second order perturbative correction to the energy is also
easily obtained as
\begin{equation}
\frac {E_2(\rho)} {N} = - \frac {1}{2 \rho} \int {d{\bf q}\over(2 \pi)^3}  
{\mid \tilde V(q) \mid^2\over\hbar^2 q^2/m} \ ,
\end{equation}
however the  resulting total energy $E(\rho)=E_1(\rho)+E_2(\rho)$ is
no longer a bound to the exact one.

If the interaction is finite the 
{\em uniform limit} approximation\cite{feenberg} may also give an 
accurate description of a dilute system. In this regime,   
correlations are assumed to be weak and 
the ground state is only slightly affected by them. This 
condition reads, in terms of the radial distribution function, as 
$\vert g(r)-1|<< 1$, as already mentioned in the 
Introduction. In the uniform limit  the HNC energy 
simplifies as:
\begin{equation}
\frac {E_{UL}(\rho)} {N} = 
\frac{1}{2} \tilde V(0) + \frac {1}{2} \int 
{d{\bf k} \over (2\pi)^3\rho}
\left [(S(k)-1) \tilde V(k) + \frac {\hbar^2
k^2}{4m} \frac {(S(k)-1)^2}{S(k)} \right ] \ ,
\label{enerULA}
\end{equation}
an expression that can be readily obtained from Eq.~(\ref{energy})
by assuming $g(r)\sim 1$, so that $\ln g(r)\approx g(r)-1$.  
The last term in the integral is the kinetic energy contribution, 
and by comparing it with (\ref{kinetic}) one readily realizes that
the momentum distribution is related to the
static structure function in the UL through the relation: 
\begin{equation}
n(k) = \frac {(S(k) -1)^2}{4 S(k)} \ .
\label{nksk}
\end{equation}
The condensate fraction, $n_0$, can then be recovered
by imposing the normalization condition (\ref{norm}). 

Minimization of $E_{UL}$ with respect to $S(k)$ gives the
Euler-Lagrange equation in the UL, 
\begin{equation}
\tilde V(k) + \frac {\hbar^2 k^2}{4 m} 
{(S^2(k) -1)\over S^2(k)} = 0 \ ,
\end{equation}
which can be solved  for the the static structure function to obtain: 
\begin{equation}
S_{UL}(k) = \frac {t(k)} {\sqrt{ t^2(k) + 2 t(k) \tilde V(k) }} \ .
\end{equation}
This expression is formally identical to the HNC one of
Eq.~(\ref{S_Euler}), where 
$V_{ph}(k)$ has been approximated by the Fourier transform 
of the bare potential $V(r)$.

We notice that $S_{UL}(k)$ 
coincides with the static sturcture function given by the RPA approximation to the
dynamic susceptibility, $\chi(k, \omega)$ \cite{kro98}, 
\begin{equation}
\chi_{RPA}(k, \omega) = \frac {\chi_0(k,\omega)}{1 - \tilde V(k)
\chi_0(k, \omega)} \ ,
\end{equation}
where $\chi_0(k,\omega)$ is the dynamic susceptibility of the free
Bose gas at zero temperature,
\begin{equation} 
\chi_0(k,\omega) = \frac {1}{\omega -t(k) +i \epsilon} - \frac
{1}{\omega+t(k) +i \epsilon} \ .
\end{equation}
The poles of $\chi_{RPA}(k,\omega)$, $\epsilon_{RPA}(k) = \left (t(k)^2+2
t(k)\tilde V(k)\right)^{1/2}$, define the excitation energies of the
system, while the imaginary part of $\chi_{RPA}(k,\omega)$ gives the
dynamic structure function, 
\begin{equation}
S_{RPA}(k,\omega) = - \frac {1}{\pi} \Im \chi_{RPA}(k,\omega)=
 \frac {t(k)}{\epsilon_{RPA}(k)}\delta(\omega - \epsilon_{RPA}(k)) \ .
\end{equation}
Integration of $S_{RPA}(k,\omega)$ over $\omega$ leads to the 
static structure function, $S_{RPA}(k)$, 
\begin{equation}
S_{RPA}(k) = {t(k) \over \sqrt{t^2(k) + 2 t(k) \tilde V(k)}} \ ,
\end{equation}
which coincides with $S_{UL}(k)$. 
 Moreover, the UL energy (\ref{enerULA}) can
be obtained by adding the RPA correction,  
$\triangle E_{RPA}$, to the uncorrelated 
energy (\ref{ener-upper}).
$\triangle E_{RPA}$ can be
evaluated by performing a coupling constant integration, 
\begin{equation}
\triangle E_{RPA} = {\frac {1}{2}}\int {d{\bf k}\over(2\pi)^3\rho}\,
\tilde V(k) \int_0^1 \left [S_{\lambda}(k) -1\right ] d\lambda \ ,
\end{equation}
$S_{\lambda}(k)$ being the RPA static structure function 
corresponding to the interaction $\lambda \tilde V(k)$. 
After a  straigthforward integration the uniform limit expression 
is recovered,
\begin{equation}
E_{UL} = \frac {1}{2} \tilde V(0) + \triangle E_{RPA} \ .
\end{equation}

A similar result has been found in Ref.\cite{sarsa02}, where the soft-sphere gas 
has been studied in the framework of the HNC theory  
with an independent pair correlated  wave function. The authors  
have shown that neglecting the {\em composite diagrams} 
and summing the {\em chain diagrams} only, the HNC/IPC approach 
leads to the RPA (and UL) energy functional (\ref{enerULA}). 
All three theories produce the same description of the soft-sphere gas 
in the low density  regime, since both uniform limit  and correlated 
theories, in conjuction with a proper minimization via the Euler 
equation, take into account the long range correlations relevant 
in this region and correctly considered by the RPA.

\subsection{Some aspects of the low density expansion }

The formalism outlined in Section II allows to describe  
the ground state of any boson liquid or gas characterized by a
hamiltonian of the form given in (\ref{hamil}).
 At very low densities, however,
all interacting Bose gases having the same scattering length 
are expected to behave in 
a similar way, as originally pointed out by Bogoliubov \cite{bogo}.  
The energy of a Bose gas follows the universal form (\ref{E_LY}) as
long as the system is dilute. However, the overall energy comes from 
a balance between kinetic and potential contributions. For hard spheres 
the energy is entirely kinetic, whereas for soft spheres 
the relative contribution of  
the potential energy may be as large as possible, according to the 
softness of the interaction.

Despite the universality exhibited by the total energy,
 it is not clear that  other ground state
properties  may show an analogous  behavior. An answer to this
question can 
be obtained by analyzing the hard and soft sphere 
gases and comparing with the predictions obtained within the 
 Bogoliubov model. 
For instance, from the Bogoliubov excitation spectrum, 
\begin{equation}
\epsilon_{B} (k)=\sqrt{t^2(k)+2 t(k) 4\pi a \hbar^2\rho/m} \ ,
\label{eps_bogo}
\end{equation}
an approximate  static structure factor can be obtained assuming 
a Feynman--like spectrum, 
\begin{equation}
S_{B}(k) = { t(k) \over \epsilon_{B}(k)} \ .
\label{sqbogo}
\end{equation}
However, $g(r)$ extracted from $S_{B}(k)$ is unphysically divergent 
at $r=0$.

In the Bogoliubov model the momentum distribution, $n_{B}(k)$, 
at $k>0$  can be written in the form\cite{huang}:
\begin{equation}
n_B(k) = {A_k^2 \over 1 - A_k^2}
\,\,\,\,\, , \,\,\,\,\,
A_k = 1 + {m\over 4\pi a \hbar^2\rho} \left[ t(k) - 
\epsilon_{B}(k) \right] \ .
\label{nkBogo}
\end{equation}
This expression coincides with $n(k)$ as obtained in the 
UL (Eq.(\ref{nksk})) when $\tilde V(k)$ is replaced by 
$4\pi a \rho \hbar^2/m$.
 
Normalization of the full momentum distribution gives the fraction 
of particles in the $k=0$ state, $n_{0B}$,
\begin{equation}
n_{0B}= 1 - {8\over 3}\sqrt{x\over \pi} \ .
\label{noBogo}
\end{equation}
The kinetic energy computed through $n_B(k)$ is divergent since 
$n_B(k)\sim k^{-4}$ at large $k$.

\section{Dilute hard spheres}

In this section variational results for 
the energy, two-body distribution function, static structure function 
and one-body density matrix (and momentum distribution)
 of a dilute gas of hard spheres are shown and discussed.  
The driving quantity of any variational approach is the total energy, 
given by the sum of the kinetic and potential terms. 
Since inside the core of the potential $g(r)=0$, 
the energy is purely kinetic and, in units of $\hbar^2/2m a^2$,  
becomes
\begin{equation}
\frac {{\bar E}(x)}{N}=\frac {\langle {\bar T} \rangle}{N}  = 
-{1\over 2} x \int d{\bf r} \, g(r) \nabla^2 \ln f(r) \ .
\label{ener-b2}
\end{equation}

We will adopt the long ranged correlation function obtained 
by solving the Euler--Lagrange equations and will analyze the 
related asymptotic behaviors. However, the HNC results for a simpler 
correlation function having a short--range structure, $f_{SR}(r)$, 
are also discussed. $f_{SR}(r)$ is choosen in such a way to minimize  
the lowest--order in the cluster expansion of the energy of the 
homogeneous gas of HS with a healing condition at a distance $d$, 
taken as a variational parameter\cite{fSR}. 
In the HS case $f_{SR}(r)$ is:
\begin{equation}
f_{SR}(r) = \left\{
\begin{array}{ll}
  0 & \,\,\,\,\, r<1 \\
  {d\over r} {\sin[K(r-1)] \over \sin[K(d-1)] } & \,\,\,\,\, r> 1 \ ,
\end{array}
\right.
\label{fr-b1}
\end{equation}
where distances are in units of $a$ and $K$ fulfills 
the equation: $cot[K(d-1)]=(Kd)^{-1}$. The latter condition  
ensures the healing properties: $f_{SR}(r\ge 1)=1$ and 
$f^\prime_{SR}(r=d)=0$.

The scaled energies per particle, ${\bar E}(x)/N$, of the HS gas 
as a function of the gas parameter, $x$, are given in 
Table~\ref{tab-1}. ${\bar E}_{EL}$ is obtained by solving the 
Euler-Lagrange equation, while in computing ${\bar E}_{SR}$  
the short--range correlation (\ref{fr-b1}) has been used. 
Both energies are computed 
within the HNC/0 approximation, justified {\em a priori} by the 
smallness of $x$, and {\em a posteriori} by the eventual agreement with the
exact DMC results\cite{boro99}, reported also in 
the Table together with the results of the LD expansion (\ref{E_LY}). 

${\bar E}_{EL}$ and ${\bar E}_{SR}$ are upper bounds to the exact 
DMC energy. Furthermore, since the solution
of the Euler--Lagrange equations yields the minimum energy for a
Jastrow type wave function, the inequality 
${\bar E}_{DMC}\leq {\bar E}_{EL}\leq {\bar E}_{SR}$ holds at each 
density. Violations of this hiearchy at low $x$--values are 
probably due to numerical inaccuracies, rather than to the HNC/0 
approximation.

\begin{figure}
\begin{center}
\includegraphics*[height=11cm]{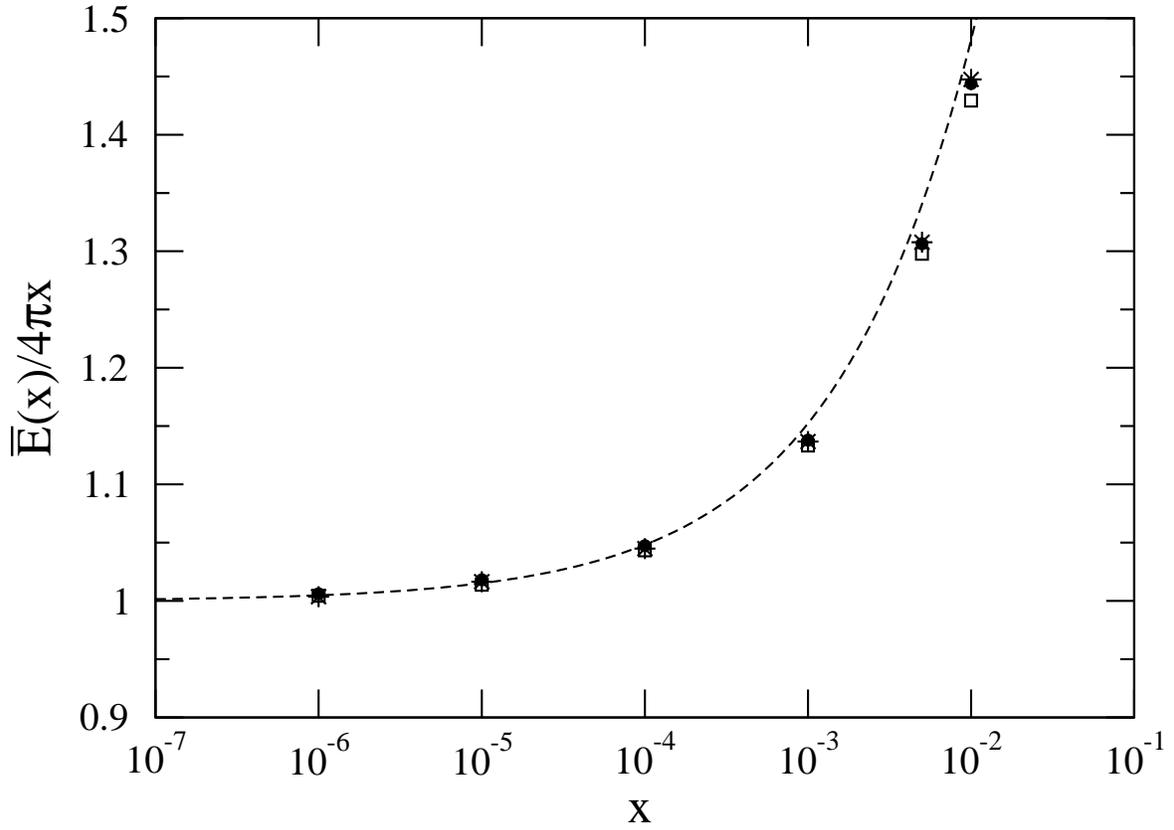}
\end{center}
\caption{Scaled energy per particle of the HS gas as a function of
  $x$. Solid circles: Euler--Lagrange results; stars: HNC/0
  results for the parametrized two--body correlation factor of
  Eq.(\ref{fr-b1}); open squares: Diffusion Monte Carlo resuls;
  dashed line: first two terms of the low density expansion,
  Eq.(\ref{E_LY}).}
\label{fig-ener}
\end{figure}

The low-density  expansion does not satisfy the upper bound property. 
However, the lowest order of the same expansion, 
${\bar E}_{LD0}/N= 4 \pi x $, is a rigorous lower bound
to the exact energy \cite{lieb1}.  The overall differences among the
EL, SR and DMC energies are very small (at most $6 \%$ in the worst case
at $x=0.1$) and both the EL and the SR results can be taken as good
estimates. Using the HNC/0 scheme is therefore well justified in the
 range of  densities explored, especially at the lowest ones. 
For the sake of comparison, 
 a Variational Monte Carlo (VMC) calculation at $x=0.1$, 
with $f_{SR}(r)$ having a healing distance  $d=6$, 
has also been carried out in order to estimate the relevance of
 the elementary diagrams. The  
result ${\bar E}_{VMC}/N = 3.74 \pm 0.02$ is to be compared with 
${\bar E}_{SR}/N = 3.97$ in HNC/0. Hence, at the highest density considered 
the elementary diagrams contribute by only $\sim 6 \%$ to the energy. 
Further differences with the DMC results have to be attributed to 
deficiencies in the two--body correlation factor   
 and to the lack of three- and higher body
correlations in the trial wave function. The energies in
these approximations are plotted in Fig.~\ref{fig-ener}, where the
subtle differences between the points are hardly appreciable, 
except at the highest $x$'s.

The influence of the optimization on the energy is rather small. 
The energy is dominated by the short range structure of the potential, 
which requires the two body distribution function to be zero inside
the hard core. The Euler--Lagrange procedure is instead important in 
establishing the long range structure of the distribution function 
$g(r)$, or, alternativeley, of the low $k$ behavior of $S(k)$.

\begin{figure}
\begin{center}
\includegraphics*[height=11cm]{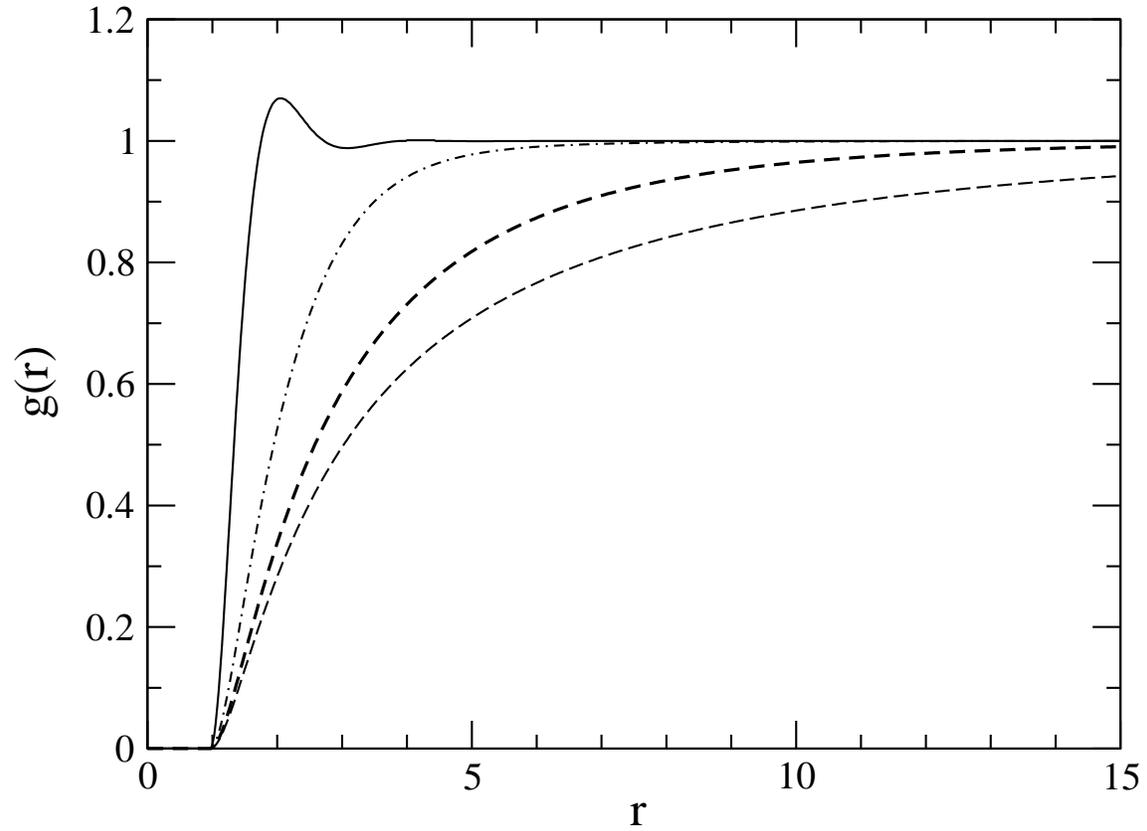}
\end{center}
\caption{EL radial distribution function $g(r)$ for HS at $x=10^{-4}$
  (short-dashed line), $x=10^{-3}$ (long-dashed line), $x=10^{-2}$
  (dot-dashed line) and $x=10^{-1}$ (solid line). 
 Distances in units of $a$. }
\label{fig-grx}
\end{figure}

The $EL$ radial distribution function is shown in Fig.~\ref{fig-grx} for several
values of the gas parameter.  At low $x$'s $g(r)$ is a
monotonically increasing function of the distance, approaching 
faster and faster  the asymptotic limit, $g(r)\to 1$, with the 
density. This behavior is readily understood recalling that $g(r)$
measures the probability of finding two particles at a distance $r$, 
and the average interatomic spacing decreases with the density.  
At large densities the radial distribution function develops a local maximum 
close to the core radius.

\begin{figure}
\begin{center}
\includegraphics*[height=11cm]{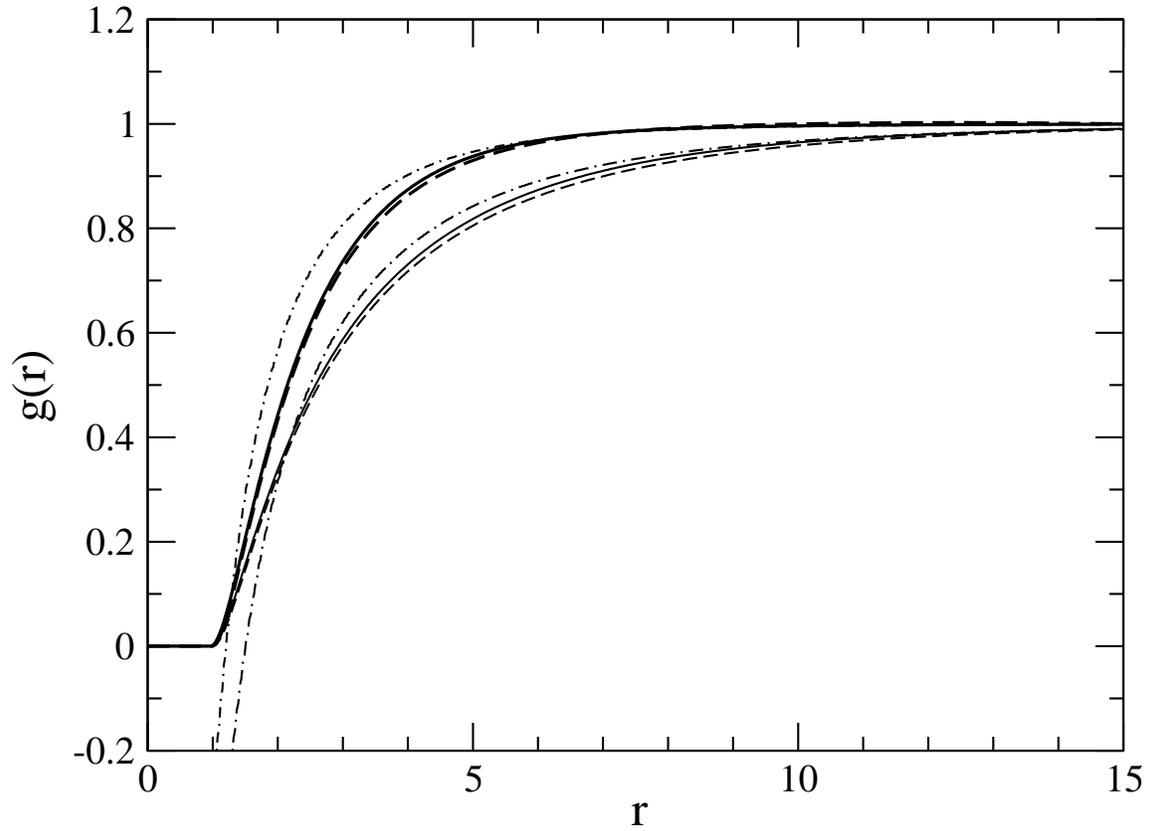}
\end{center}
\caption{EL (solid lines), SR (dashed lines) and Bogoliubov 
  (dash--dotted lines) radial distribution
  functions for HS. Upper and lower curves correspond to $x=0.005$ and
  $x=0.001$, respectively. Distances in units of $a$.}
\label{fig-grcompx}
\end{figure}

The radial distribution functions, $g_{EL}(r)$, solutions of the EL 
equation at $x=0.001$ and $x=0.005$, are compared in
Fig.~\ref{fig-grcompx} with the corresponding short-range  ones, $g_{SR}(r)$,
computed from $f_{SR}(r)$.  At these densities, the energy of the 
system is accurately described by the 
expansion of Eq.~(\ref{E_LY}), and it is natural to 
compare other ground state quantities to those corresponding to
the Bogoliubov approximation.
Fig.~\ref{fig-grcompx} also shows the radial distribution function, 
$g_{B}(r)$, obtained as the Fourier transform of $S_{B}(k)$, 
\begin{equation}
S_{B}(k) = { k^2 \over \sqrt{ k^4 + 16\pi x k^2 }} \ .
\label{SqBogo-b1}
\end{equation}

\begin{figure}
\begin{center}
\includegraphics*[height=11cm]{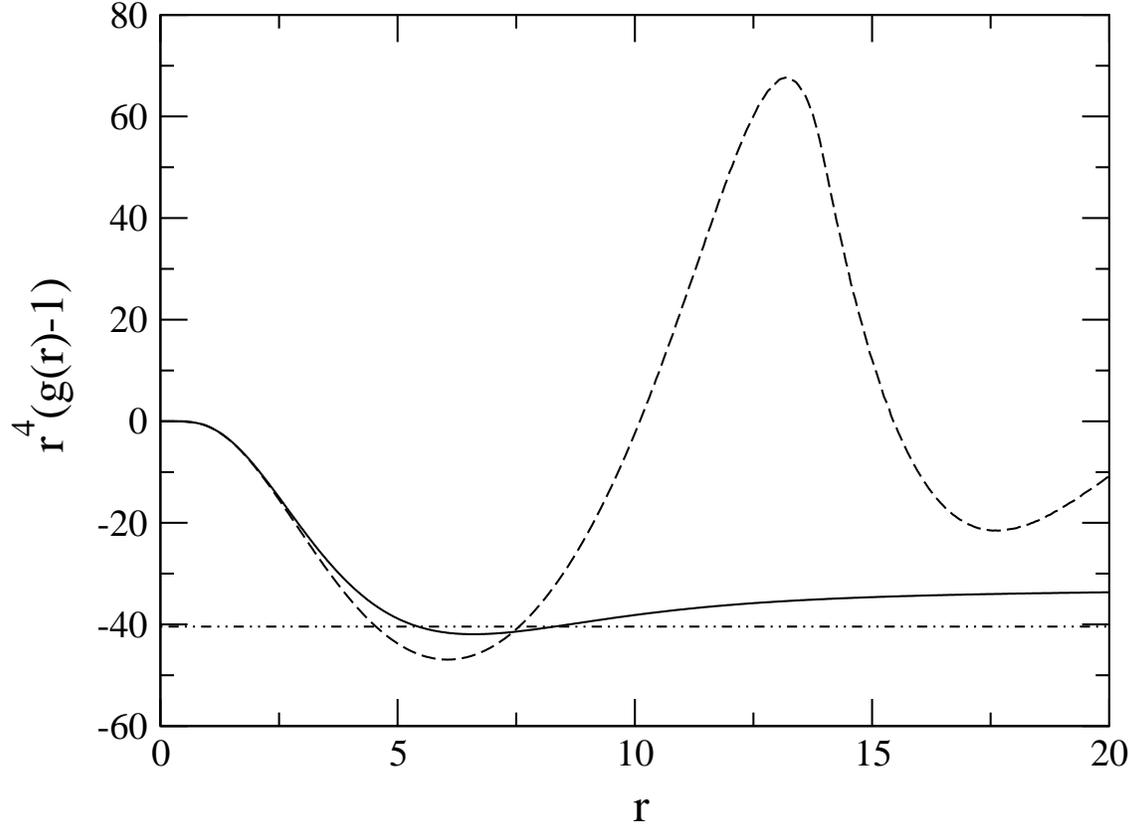}
\end{center}
\caption{Long range  structure of the HS $g(r)$ at $x=0.005$ in the EL 
(solid line),  SR (dashed line) cases compared with the $r\rightarrow \infty$
contribution in the  Bogoliubov approximation (dot--dashed line).
 Distances in units of $a$. }
\label{fig-r4gm1}
\end{figure}

$g_{EL}(r)$ and $g_{SR}(r)$ are close at short distances: 
they both vanish inside the core and approach similarly the 
unity. However, differences with $g_{B}(r)$
are significant. $g_{B}(r)$ becomes unphysically
negative at short distances to finally diverge at $r\to 0$. In fact,
$S_{B}(k)$ realistically reproduces only the low $k$ behavior 
of the static structure function in dilute systems. 
This corresponds to the large--$r$ region in the associated radial distribution function.  
For the same reason, $g_{B}(r)$ never develops a maximum. 
The three radial distribution functions have different asymptotic 
behaviors, as shown in Fig.~\ref{fig-r4gm1}, where $r^4(g(r)-1)$ 
is given in the EL, SR and Bogoliubov approaches at $x=0.005$. 
Both $g_{EL}(r)-1$ and $g_{B}(r)-1$ behave as
$r^{-4}$ at large $r$, a property not shown by
$g_{SR}(r)-1$ since $f_{SR} (r)$ does not have the
appropriate long range behavior, $f(r\to\infty)-1\sim r^{-2}$.  
Actually, the long range behavior of $g(r)$ is related to the sound 
velocity, $c$ (in units of $\hbar/2ma$), by\cite{feenberg} 
\begin{equation}
g(r \to \infty ) \sim 1 - \frac {1}{\pi^2 c x }
\frac {1} {r^4} \ ,
\end{equation}
and consequently, the static sturcture function goes like
\begin{equation}
S(k\to 0) \to {k \over c} \ .
\label{lowSbehaviour}
\end{equation}
In the Bogoliubov approximation,
\begin{equation}
g_{B}(r \to \infty) \sim  1 - \frac {1}{4 \pi^{5/2} x^{3/2} } \frac {1}{r^4} \ ,
\end{equation}
and $c_{B} = \sqrt{16 \pi x}$,
consistent with the value provided by the compressibility, $\kappa_T$, 
$c=(\rho m \kappa_T)^{-1/2} /(\hbar/2ma)$, 
by keeping only the first term in the low-density expansion of the energy. 
At $x=0.005$ the constant value of 
$r^4(g_{B}(r)-1)=-40.4$ gives $c_{B}=0.50$, smaller 
than the estimated $c_{EL}=0.61$.

\begin{figure}
\begin{center}
\includegraphics*[height=11cm]{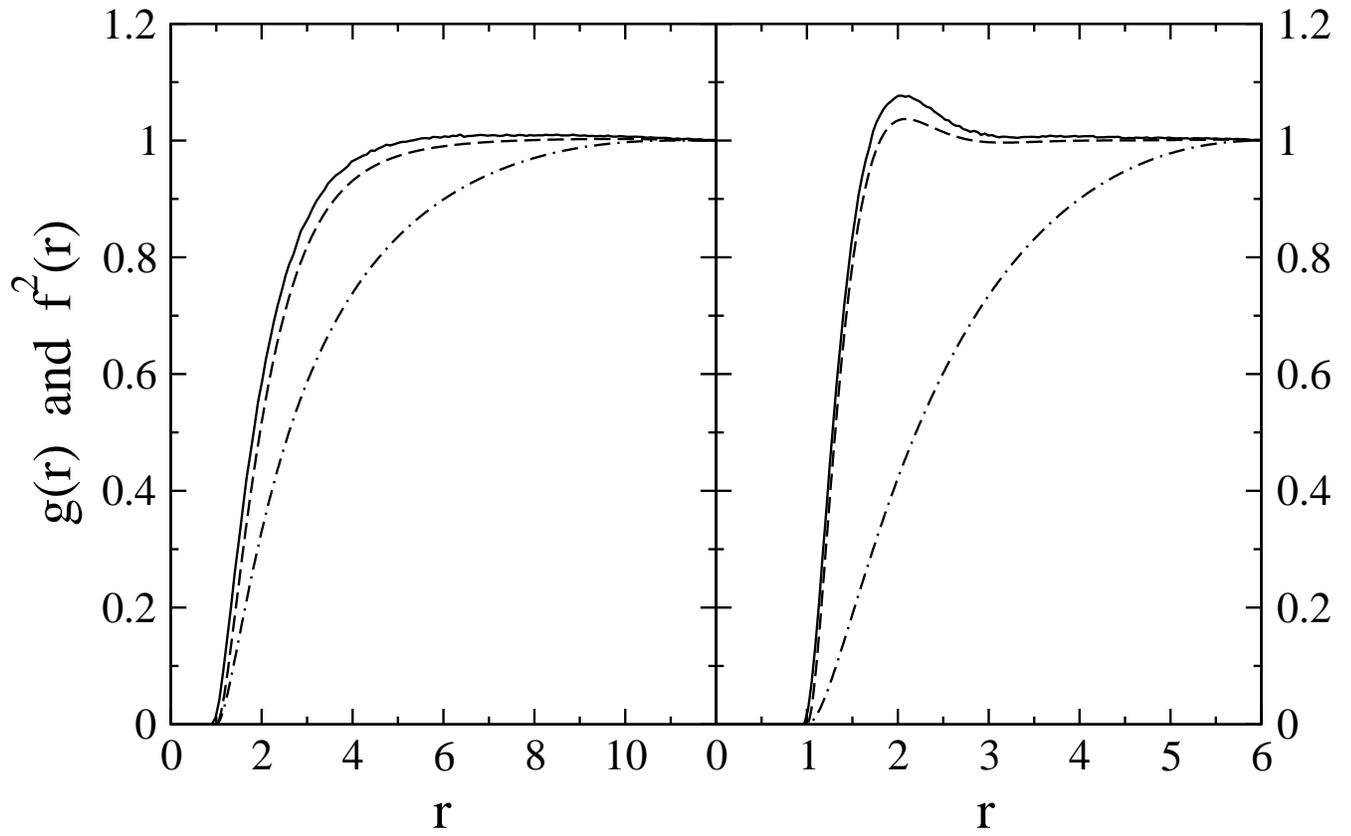}
\end{center}
\caption{
$g_{VMC}(r)$ (solid line) and $g_{HNC/0}(r)$ 
(dashed line) computed from  $f^2_{SR}(r)$ of Eq.(\ref{fr-b1}) 
(dot--dashed line) for HS at $x=0.01$ (left panel) 
and $x=0.1$ (right panel). Distances in units of $a$. 
}
\label{fig-gSSVMC}
\end{figure}

The short-range  radial distribution function is compared in Fig.~\ref{fig-gSSVMC} 
with the VMC one at $x=0.01$ 
and $0.1$, for the same two--body correlation
factor. We notice two aspects: ($i$) the
inclusion of the elementary diagrams enhances the peak of the radial distribution
function, 
and ($ii$) there is a remarkable difference between 
$f_{SR}^2(r)$ and $g(r)$. The many-body contributions included by the 
HNC scheme cannot be neglected and, therefore, $g(r)$ cannot be 
approximated by its lowest order cluster expansion value, 
$g_{LO}(r)=f^2(r)$.

\begin{figure}
\begin{center}
\includegraphics*[height=11cm]{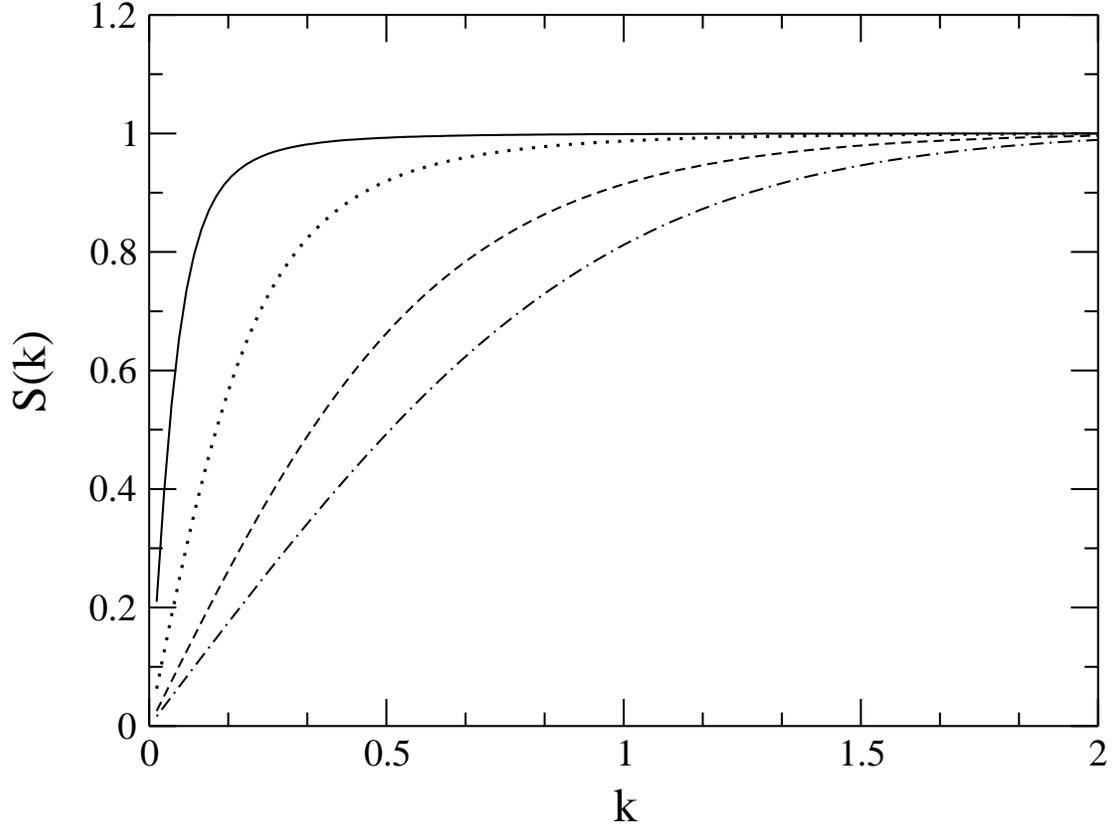}
\end{center}
\caption{HS EL static structure factor $S(k)$ at $x=10^{-4}$
(solid line), $x=10^{-3}$ (dotted line), $x=5\cdot 10^{-3}$ (dashed
line) and $x=10^{-2}$ (dot--dashed line). 
Momenta in units of $a^{-1}$.
}
\label{fig-Sk}
\end{figure}

\begin{figure}
\begin{center}
\includegraphics*[height=11cm]{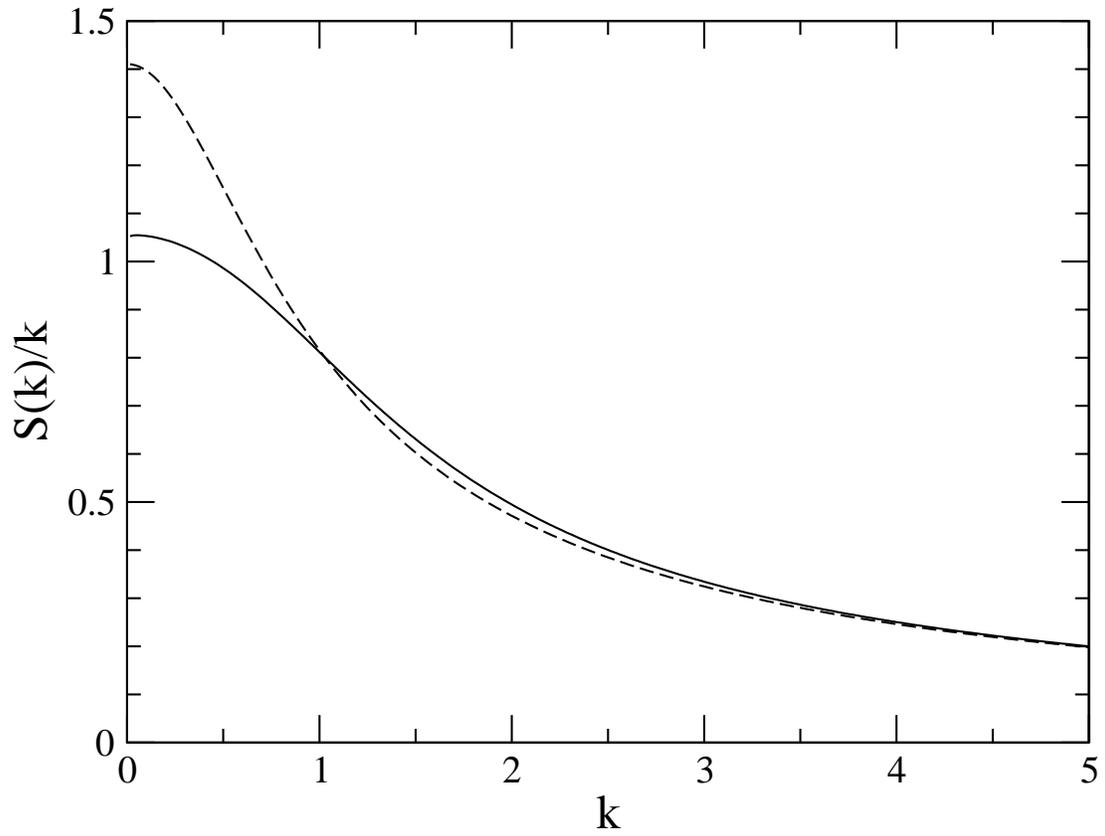}
\end{center}
\caption{HS $S(k)/k$ in the EL (solid line) and Bogoliubov 
  (dashed line) cases at $x=0.01$.
Momenta in units of $a^{-1}$. 
}
\label{fig-Sqoverq}
\end{figure}

The optimal structure function $S(k)$ is shown for several 
values of $x$ in Fig.~\ref{fig-Sk}. The slope of $S(k)$
becomes smaller when $x$ increases because the speed of sound
increases with the density. Furthermore, the linear low 
$k$ behavior, which guarantees the correct low energy 
excitation spectrum, is evident. However, the linearity 
of the static structure holds only at {\em very} low $k$, as can be 
seen from Fig.~\ref{fig-Sqoverq}, where the ratio
$S(k)/k$ is shown in both the EL and the Bogoliubov cases at
$x=0.01$. In the later case the linear regime is valid only when
$k\ll\sqrt{16\pi x}=c_{B}$ (see Eq.(\ref{SqBogo-b1})).

\begin{figure}
\begin{center}
\includegraphics*[height=10.5cm]{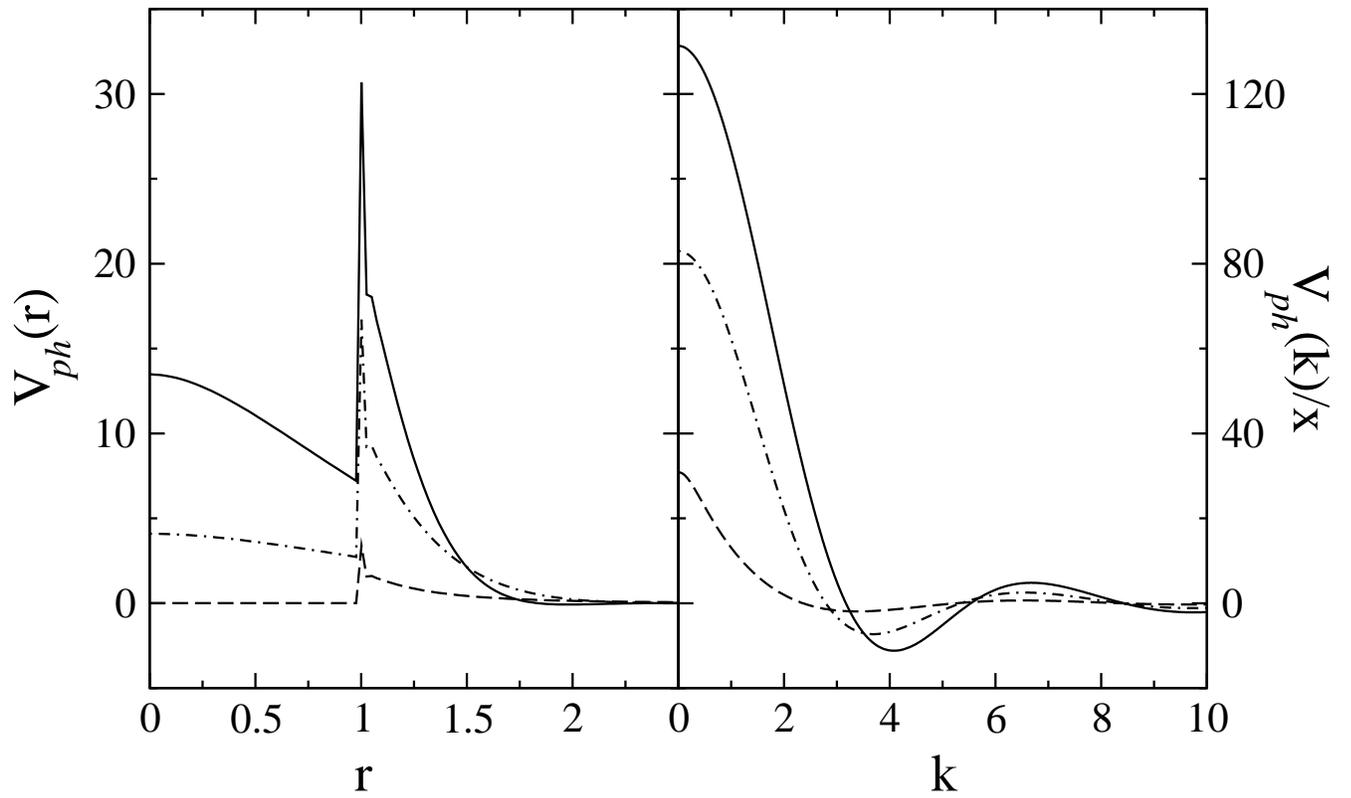}
\end{center}
\caption{EL Particle--Hole interaction of the HS gas 
at $x=0.1$ (solid line), $x=0.05$ (dot-dashed line) 
and $x=0.001$ (dashed line). 
Distances in units of $a$ and momenta in units of $a^{-1}$. 
}
\label{fig-VphHS}
\end{figure}

The particle-hole interaction $V_{ph}$ (Eq.~(\ref{Vph})), 
is shown in Fig.~\ref{fig-VphHS}
for $x=0.001$, $x=0.05$ and $x=0.1$. The left panel gives 
$V_{ph}$ in $r$--space. Two regions are separated by a 
discontinuity at $r=1$, produced by the term 
$\mid\nabla\sqrt{g(r)}\mid^2$ in Eq.~(\ref{Vph}) since the 
first derivative of the HS radial distribution function 
is discontinuous at  the core. As already noticed, 
 $-V_{ph}(r<1)$ coincides with the induced interaction,
$\omega_I(r)$ (\ref{Induced}). 
 At the lowest densities
$\mid\nabla\sqrt{g(r)}\mid^2$ dominates and the strength of
$V_{ph}(r)$ is almost entirely exhausted by it. 
The right panel displays the Fourier transform of $V_{ph}$ at the same densities.  
$V_{ph}(k)$ is an oscillating function of $k$, with its  
highest amplitude at $k=0$, and $V_{ph}(k=0)=c^2/2$. Notice that 
the figure shows $V_{ph}(k)/x$, therefore $V_{ph,B}(k)/x$ would be
$8 \pi$ constant and independent of $k$ and $x$, which turns out 
to be a very bad approximation to $V_{ph}(k)/x$. However calculating 
the speed of sound from the two terms of Eq. (\ref{E_LY}), one gets
$c^2/2x=8\pi+128 \pi^{1/2} x^{1/2}$, in much better agreement with
EL $V_{ph}(k=0)/x$.

\begin{figure}
\begin{center}
\includegraphics*[height=10.5cm]{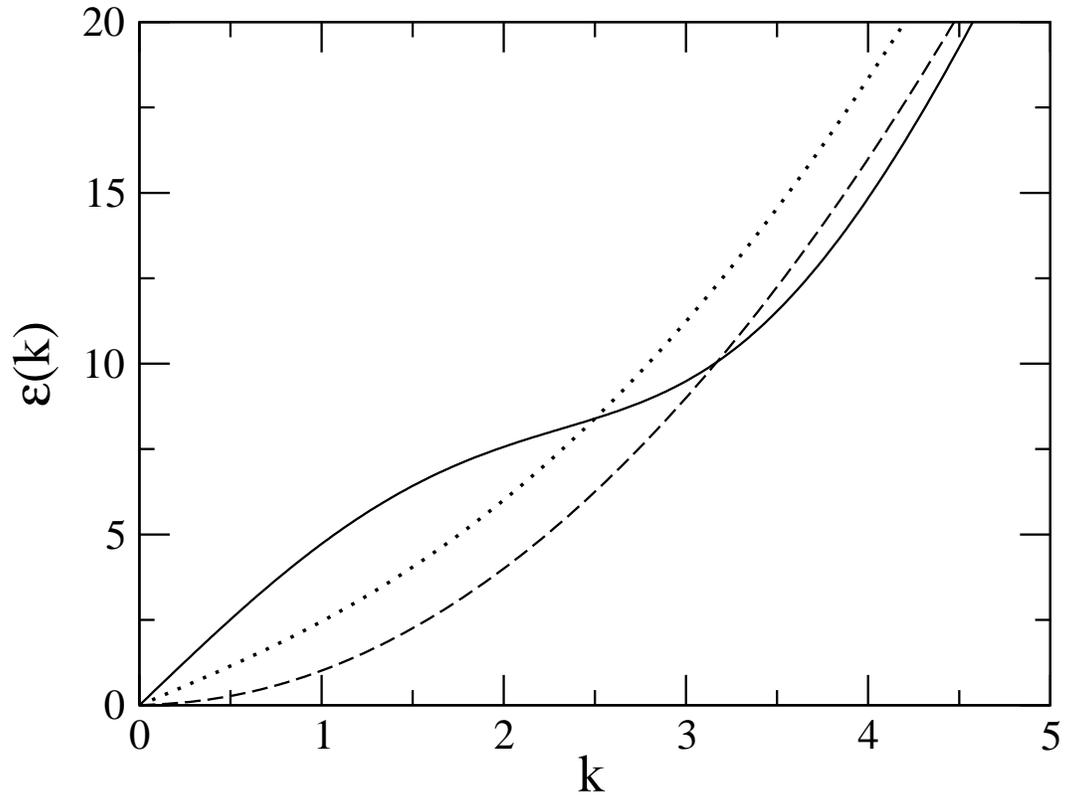}
\end{center}
\caption{Excitation spectrum of the HS gas.  
Solid line: EL at $x=0.1$, dotted line: Bogoliubov at $x=0.1$, dashed
  line: EL at $x=0.001$. 
Momenta in units of $a^{-1}$. 
}
\label{fig-eqHS}
\end{figure}

The Bogoliubov  estimate of the particle--hole interaction 
 (Eqs.(\ref{S_Euler}) and 
(\ref{SqBogo-b1})) is $V_{ph,B}(k)=8\pi x$ and it 
is a rather poor approximation to $V_{ph}(k)$.
In this case, the  excitation spectrum, 
$\epsilon(k)=t(k)/S(k)\equiv \sqrt{k^4 + 2 k^2 V_{ph}(k)}$, 
does never develop a rotonic structure,  
 since $V_{ph,B}(k)$ is a constant. 
In the EL approach, instead, the first oscillations of 
$V_{ph}(k)$ (and of the static structure function) are large enough to produce a 
maxon--roton behavior at the largest $x$ value 
($x=0.1$). The two spectra are explicitely shown in 
Fig.~\ref{fig-eqHS}. The Figure also gives $\epsilon(k)$ at 
$x=0.001$, where the differences between the EL and Bogoliubov 
results can hardly be resolved.

The last quantity analyzed is the  momentum distribution, $n(k)$.  
Table~\ref{tab-2} reports the condensate
fraction, $n_0$, and the kinetic energy for both the EL and 
the SR correlation factors. The Table gives the kinetic 
energies estimated by Eq.~(\ref{kinetic}), $T_n$, and 
by the second term of Eq.~(\ref{energy}), $T_g$. At low 
$x$ values the two estimates almost coincide, whereas, 
discrepancies at larger densities are to be ascribed to the 
lack of the elementary diagrams contribution in the cluster 
expansions of the radial distribution function and the momentum distribution.
 It has to be noticed that 
these diagrams play different roles in the two 
expansions\cite{fantoni78}.

The condensate fraction decreases with $x$, running from 
$\sim 95 \%$ at $x=0.001$ to $\sim 33 \%$ for $x=0.08$. 
The corresponding predictions for $n_{0B}$ (Eq.\ref{noBogo}) 
are $n_{0B}(x=0.001)=0.95$ and $n_{0B}(x=0.08)=0.57$. 
This points again to the failure of the Bogoliubov model at large 
densities. 

\begin{figure}
\begin{center}
\includegraphics*[height=10.5cm]{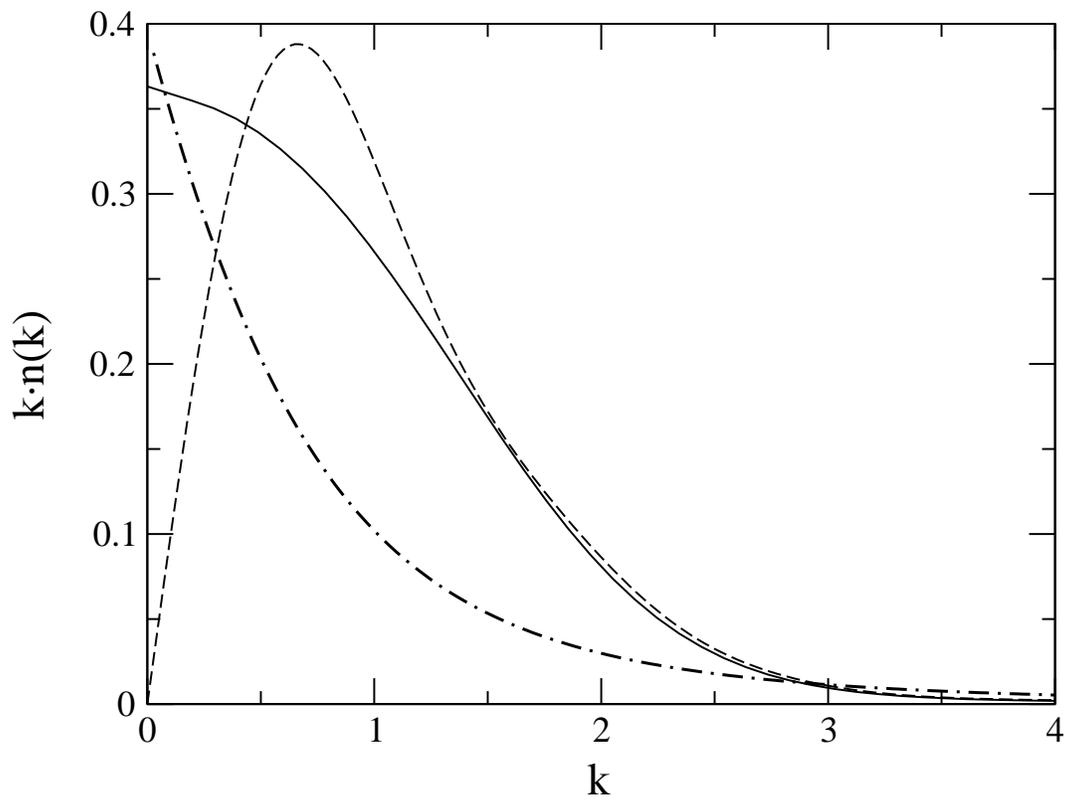}
\end{center}
\caption{HS momentum distribution at $x=0.05$.
  Solid line: EL; dashed line: SR; dot--dashed line: Bogoliubov.
Momenta in units of $a^{-1}$. 
}
\label{fig-nkcomp}
\end{figure}

The momentum distributions in the EL, SR and 
Bogoliubov cases at $x=0.05$ are shown in Fig.~\ref{fig-nkcomp}. 
The optimal $n(k)$ has the long--wavelength 
limit\cite{Gav64}
\begin{equation}
\lim_{k \rightarrow 0 } k\,n(k) = \frac {n_0 c}{4} \ ,
\label{knk0EL}
\end{equation}
while $n_B(k)$ satisfies an analogous relation, with $n_0=1$ and $c=c_B$. 
In contrast, $n_{SR}(k)$ does not behave as $1/k$ at 
$k\rightarrow 0$, and $k\,n(k)$ vanishes at the origin.
This fact is due to the lack of the proper long range structure 
in $f_{SR}$.

\begin{figure}
\begin{center}
\includegraphics*[height=10.5cm]{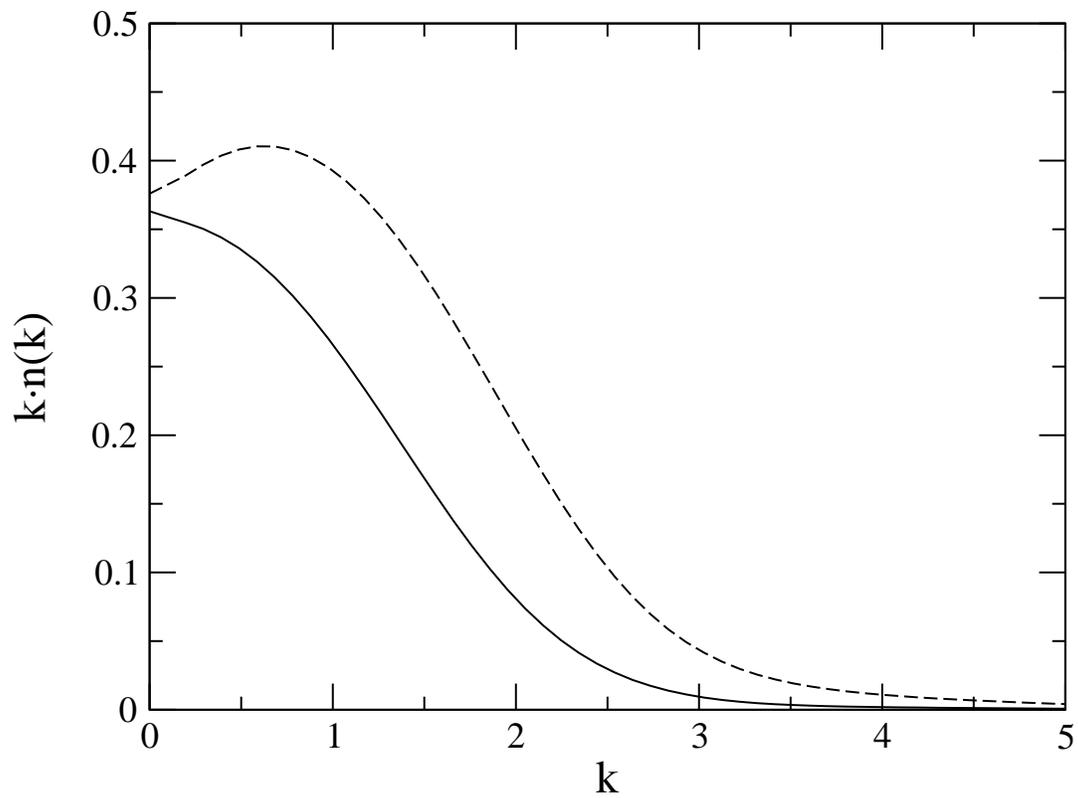}
\end{center}
\caption{HS EL momentum distributions at $x=0.05$ (solid line) 
 and $x=0.08$ (dashed line).
Momenta in units of $a^{-1}$. 
}
\label{fig-nkHS-EL}
\end{figure}

Fig.~\ref{fig-nkHS-EL} shows the optimal momentum distribution at
$x=0.08$ and 0.05.  At the lowest $x$ values 
$k\,n(k)$ is a monotonically decreasing function of $k$.  
Actually, this is also the case in the Bogoliubov approximation 
at any density. When $x$ increases, $k\,n(k)$ develops a peak 
in the EL case at low $k$. This  maximum can be
considered as a genuine effect of the short range  correlations in the EL 
approach.
Notice that the value at the origin of
$k\,n(k)$ results from a competition between $n_0$ and $c$. 
These two quantities largely vary with the density 
($n_0$ decreases and $c$ increases with increasing 
density), but the overall variation of 
$\lim_{k \rightarrow 0 }k\,n(k)$ is less pronounced.

\section{Dilute soft spheres}

The soft spheres potential (Eq.~(\ref{vpotss})) is characterized by a
core radius $R$ and a potential height $V_0$ that determine the
scattering length $a$ according to Eq.~(\ref{length}).  
The SS scattering length is always smaller
than the core radius, approaching it when $V_0$ increases. 
 In the very dilute regime,  HS and SS systems,  
having the same scattering length, 
are expected to have the same energy, while their
separate kinetic and potential contributions may differ. 
Moreover, at low $x$--values  the total energy is
well reproduced by the two terms of the low density expansion of
Eq.~(\ref{E_LY}). In this section, we will study the deviations from 
this behavior and the influence of the shape of the potential on the 
energy and its components. To this aim, two SS potentials with the same 
scattering length and different radii are considered: the SS10 and 
SS5 potentials, having $R=10$ and $R=5$ in units of $a$, respectively. 
The corresponding heights are $V^{SS10}_0=0.00681670$ and 
$V_0^{SS5}=0.06308561$ in units of $\hbar^2/2ma^2$.

Table~\ref{tab-3} reports the scaled total, kinetic and potential 
energies per particle for SS10 and SS5 in the EL, SR and UL
cases, compared with the available exact DMC results
\cite{boro99} at $x=0.0001$ and $x=0.01$. The low-density 
expansion yields ${\bar E}_{LD}/N(x=0.0001)=0.001317$ and
${\bar E}_{LD}/N(x=0.01)=0.1862$. 
The lower bound energies are 
${\bar E}_{lb}(x=0.0001)=0.001257$ and ${\bar E}_{lb}(x=0.01)=0.1257$. 
The upper bound energies (\ref{ener-upper}) depend on the shape of the 
potential, and, for SS10 and SS10, give:
${\bar E}_{ub}^{SS10}/N(x=0.0001)=0.0014277$, 
${\bar E}_{ub}^{SS5}/N(x=0.0001)=0.0016516$, 
${\bar E}_{ub}^{SS10}/N(x=0.01)=0.14277$, and
${\bar E}_{ub}^{SS5}/N(x=0.01)=0.16516$. 
Both, the upper and the lower bound properties are  here satisfied. 
The shape dependence is weaker at low $x$, in fact the difference 
between the $SS10$ and $SS5$ energies is smaller at $x=0.0001$ 
than at $x=0.01$. 
On the other hand, the kinetic and potential energies exhibit
 a potential dependence at any 
$x$--value. The  IPC results of Ref.\cite{sarsa02} are also given in 
the Table.  
The EL energies are very close to the DMC ones at $x=0.0001$ and 
the variational hiearchy is always fulfilled (DMC$<$EL$<$SR). 
As expected, the UL works better for the weaker SS10 potential, 
as well as the IPC wave function. The Jastrow wave function is 
variationally preferred to the IPC one at any $x$, as shown 
by the comparison between the EL and IPC energies. The good 
agreement between ${\bar T}_g$ and ${\bar T}_n$ seems to show that 
the HNC/0 approximation in the EL approach does not invalidate this 
conclusion.
  
\begin{figure}
\begin{center}
\includegraphics*[height=10.5cm]{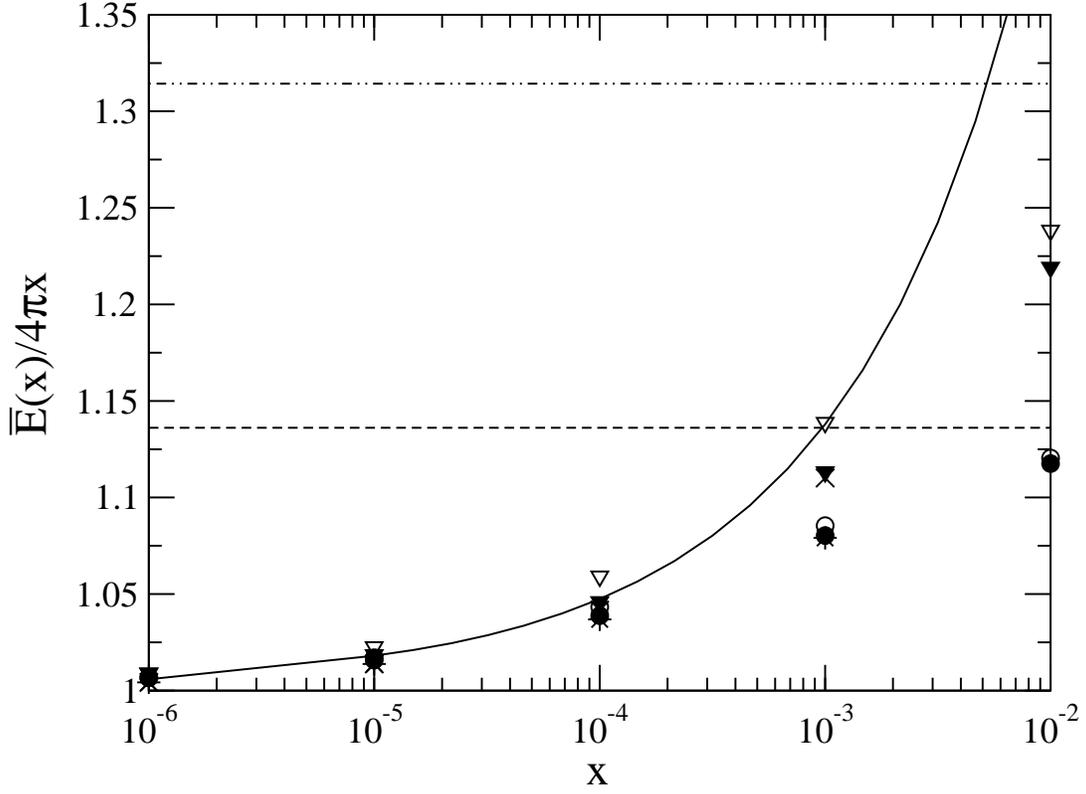}
\end{center}
\caption{Scaled energy per particle for the SS5 (triangles) and SS10 
(circles) potentials in the EL (filled triangles and circles) and  
IPC (empty triangles and circles). The stars and crosses are the 
available DMC energies for SS10 and SS5, respectively. The  solid 
line represents   the EL energies for the HS potential. The horizontal 
lines give the upper bounds for the SS5 (dash--double dotted) and 
SS10 (dashed) potentials. In these units, the common lower bound 
equals 1. 
} 
\label{fig.SS}
\end{figure}

The behavior of the scaled energy per particle along $x$ is shown 
in Fig.~\ref{fig.SS}, in units of $4\pi x$, for 
the SS10, SS5 and HS potentials, in the EL, IPC and DMC cases, 
and compared with the upper bounds. The lower bound equals 
to unity in this units.  
Starting at  $x\geq 10^{-3}$,  clearly appears a
shape dependence of the energy with the {\em harder} SS5 potential 
energies closer to the HS ones. Moreover, it is also to be stressed 
a dependence on the quality of the wave function 
in the same region, since 
differences between the Jastrow and IPC cases become evident 
for SS5. The quality of the upper bound improves 
when $R$ increases, because $V_0$ decreases at fixed $a$ and 
the perturbative expansion is expected to converge faster. 

The energy and, in general, the structure of the ground state 
depends, to a large extent, on the two--body correlation
factor employed. In the EL and UL cases, $f(r)$ is a derived
quantity as the optimization is accomplished by varying 
$g(r)$ or $S(k)$. In the SR case, $f_{SR}(r)$ is an input function 
containing some variational parameters. In the present work   
 for the SS gas we employ 
\begin{equation}
f_{SR}(r) = 1- b e^{-c r^2} \ ,
\label{fp-SS}
\end{equation}
which is flexible enough to obtain a reasonable value of the
energy. Notice that the SS potential does  not require 
the correlation function to be zero at  the origin, so $b\neq 1$.

\begin{figure}
\begin{center}
\includegraphics*[height=11cm]{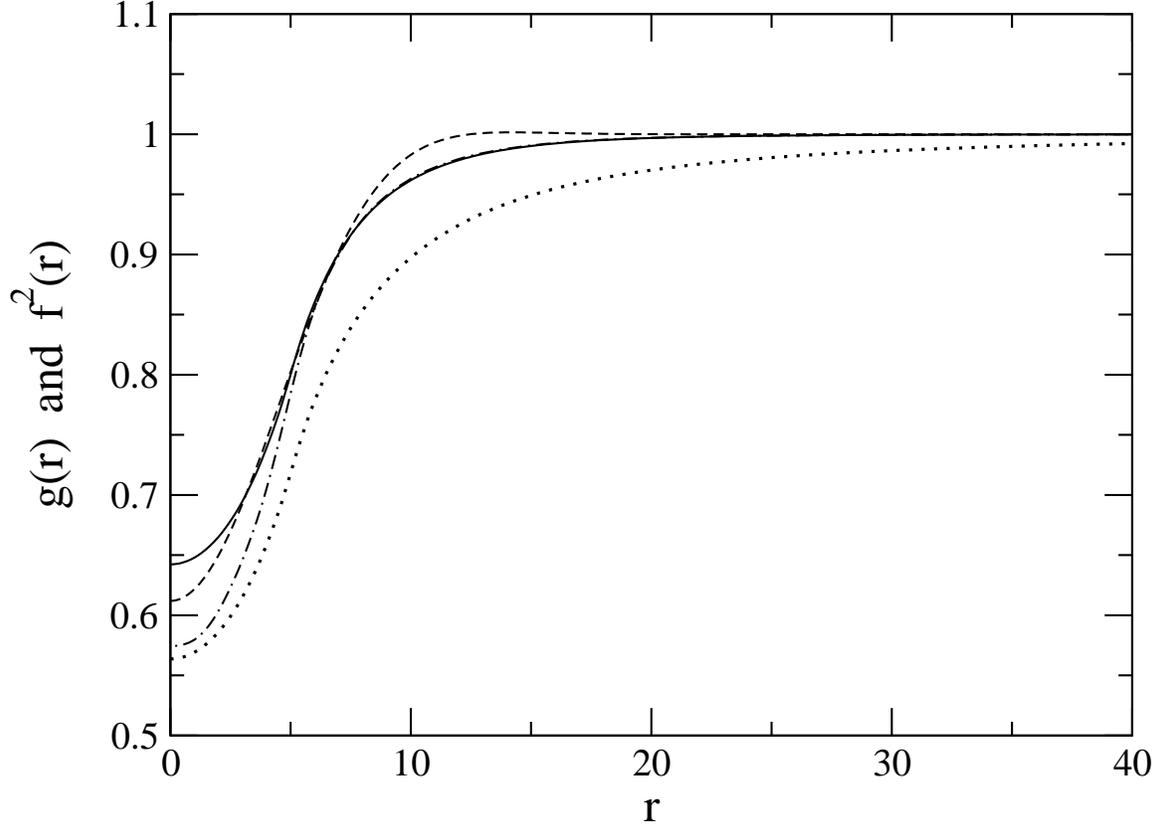}
\end{center}
\caption{ Two-body distribution functions for the SS5 potential 
at $x=0.001$. 
The solid, dot--dashed and dashed lines correspond to the  
EL, UL and SR cases, respectively. The two--body correlation
function, $f^2(r)$, as extracted from the EL $g(r)$ is 
plotted as a dotted line.  
Distances in units of $a$. 
}
\label{fig-gr}
\end{figure}

 Figure~\ref{fig-gr} shows $g(r)$ for SS5 at $x=0.001$ 
in the EL, UL and SR cases and $f_{EL}^2(r)$. Many--body 
effects make $g_{EL}(r)$ quite different from $f_{EL}^2(r)$, 
both at low and intermediate $r$--values. The EL radial distribution function  
is softer at the origin, and, in general, 
less repulsive than the short-range  one. The   radial distribution function in
the uniform limit approximation is 
close to $f_{EL}^2$ at small distances, approaching $g_{EL}(r)$
 at $r\sim 5$.  The radial distribution function  obtained by the Bogoliubov
approach (not shown in the figure) would exhibit, as for the HS
case, 
an unphysically divergent behavior at short distances.

\begin{figure}
\begin{center}
\includegraphics*[height=11cm]{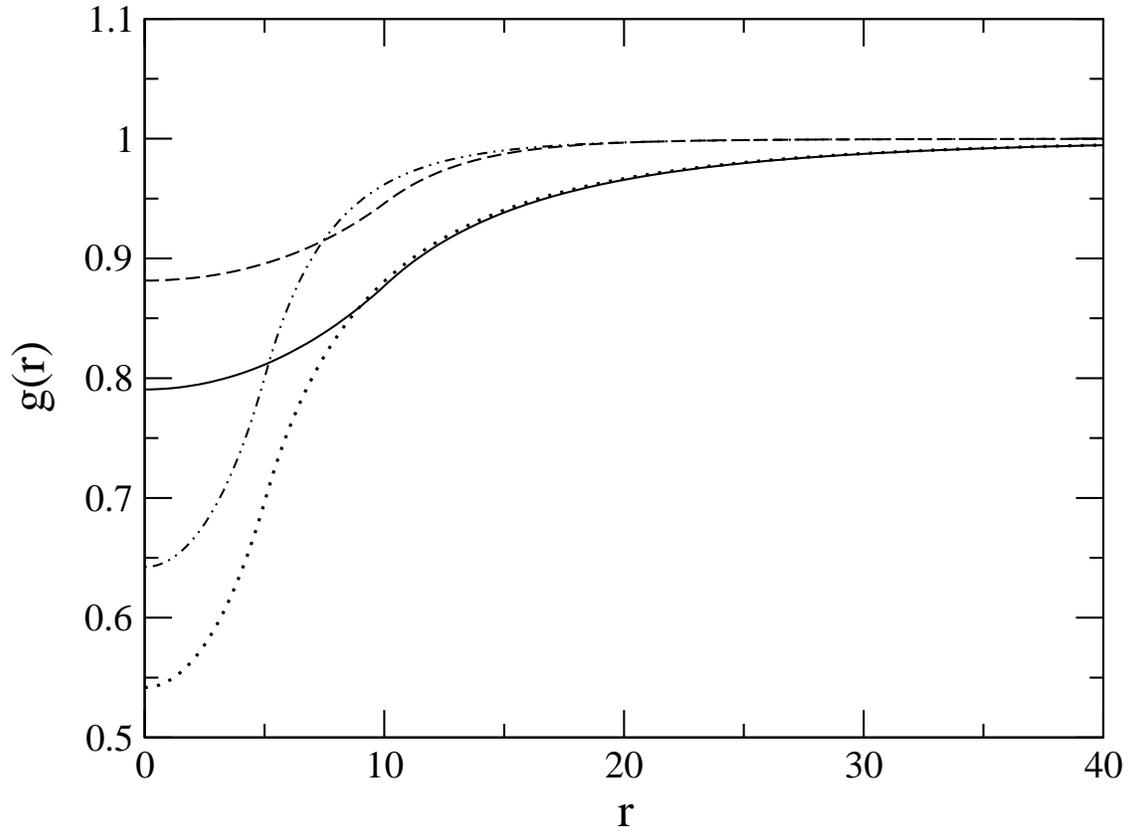}
\end{center}
\caption{EL radial distribution  functions 
for SS10 and SS5 at 
$x=0.0001$ (solid line for SS10 and dotted line for SS5) and 
$x=0.001$ (dashed line for SS10 and dashed--dot--dot line for SS5). 
 Distances in units of $a$. 
}
\label{fig-grSSa}
\end{figure}

The dependence of the radial distribution function on the shape of
the potential and on $x$ is illustrated in Fig.~\ref{fig-grSSa},
 containing the EL distribution functions for SS5 and SS10 
at $x=10^{-4}$ and $x=10^{-3}$. The central hole of the radial distribution function 
is obviously deeper for the more repulsive SS5 potential. 
It becomes less pronounced at higher $x$, since the 
probability of finding two bosons at short relative distances 
increases with the density. The universal behavior in $x$ is 
recovered at large distances, where the Bogoliubov 
approach becomes reliable. 

\begin{figure}
\begin{center}
\includegraphics*[height=11cm]{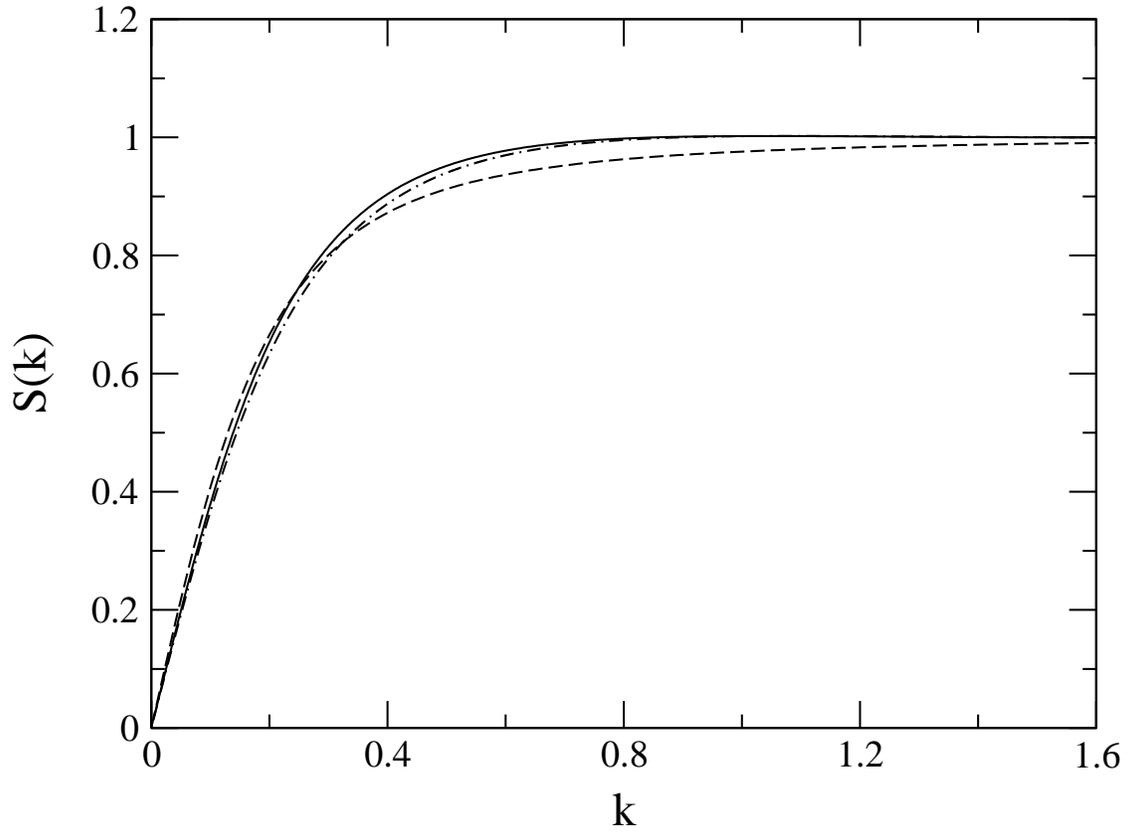}
\end{center}
\caption{The SS static structure function $S(k)$ at $x=0.001$ 
for SS5. The solid, dashed and dot--dashed lines 
stand for the EL, Bogoliubov and the UL cases, respectively. 
Momenta in units of $a^{-1}$. 
}
\label{fig-sssx0}
\end{figure}

\begin{figure}
\begin{center}
\includegraphics*[height=11cm]{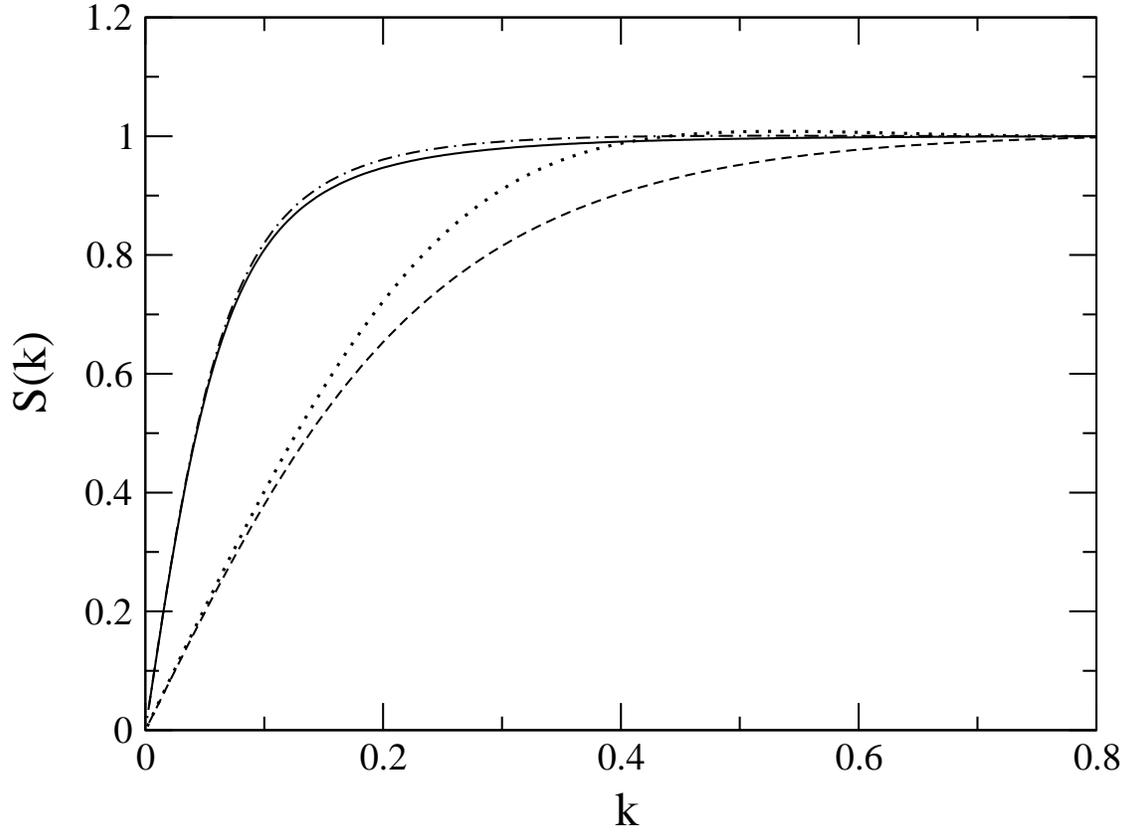}
\end{center}
\caption{EL static structure functions  
for SS10 and SS5 at 
$x=0.0001$ ( solid line for SS5 and dot--dashed line for SS10),
 and $x=0.001$ (dashed line for SS5 and dotted line for SS10). 
Momenta in units of $a^{-1}$. 
}
\label{fig-sss2x}
\end{figure}

The static structure functions, in the EL, UL and Bogoliubov 
approaches, are plotted in Fig.~\ref{fig-sssx0} for the SS5 
interaction. In all cases, $S(k)$ grows linearly at the origin, 
and the slope, governed by the sound velocity
(Eq.~(\ref{lowSbehaviour})),  is similar in all three cases.
The Bogoliubov sound velocity,
$c_{B}=\sqrt{16\pi x}$, 
is smaller than the UL estimate, 
\begin{equation}
c_{UL} = \sqrt{\frac{8}{3} \pi R^3 x V_0} \ ,
\label{eq-csound}
\end{equation}
and consequently, the slope at low momenta in $S_{B}(k)$ is 
slightly larger 
than in $S_{UL}(k)$. $S_{UL}(k)$ 
approaches $S_{EL}(k)$ and the asymptotic value 
faster than $S_{B}(k)$. Actually, at a given density 
$S_{B}(k)$ and $c_{B}$ are identical for all the SS interactions 
with the same scattering length. The short-range  $g(r)$  does not have the
proper asymptotic behavior and produces a static structure function that does not
vanish at the origin and does not increase linearly at low $k$.
 $S_{EL}(k)$ at $x=0.0001$ and $x=0.001$ are shown in
Fig.~\ref{fig-sss2x} for SS5 and SS10. 
The SS10 sound velocity is smaller and 
the slope of $S(k)$ at the origin is larger.  
The differences are more evident at the largest $x$. 

\begin{figure}
\begin{center}
\includegraphics*[height=11cm]{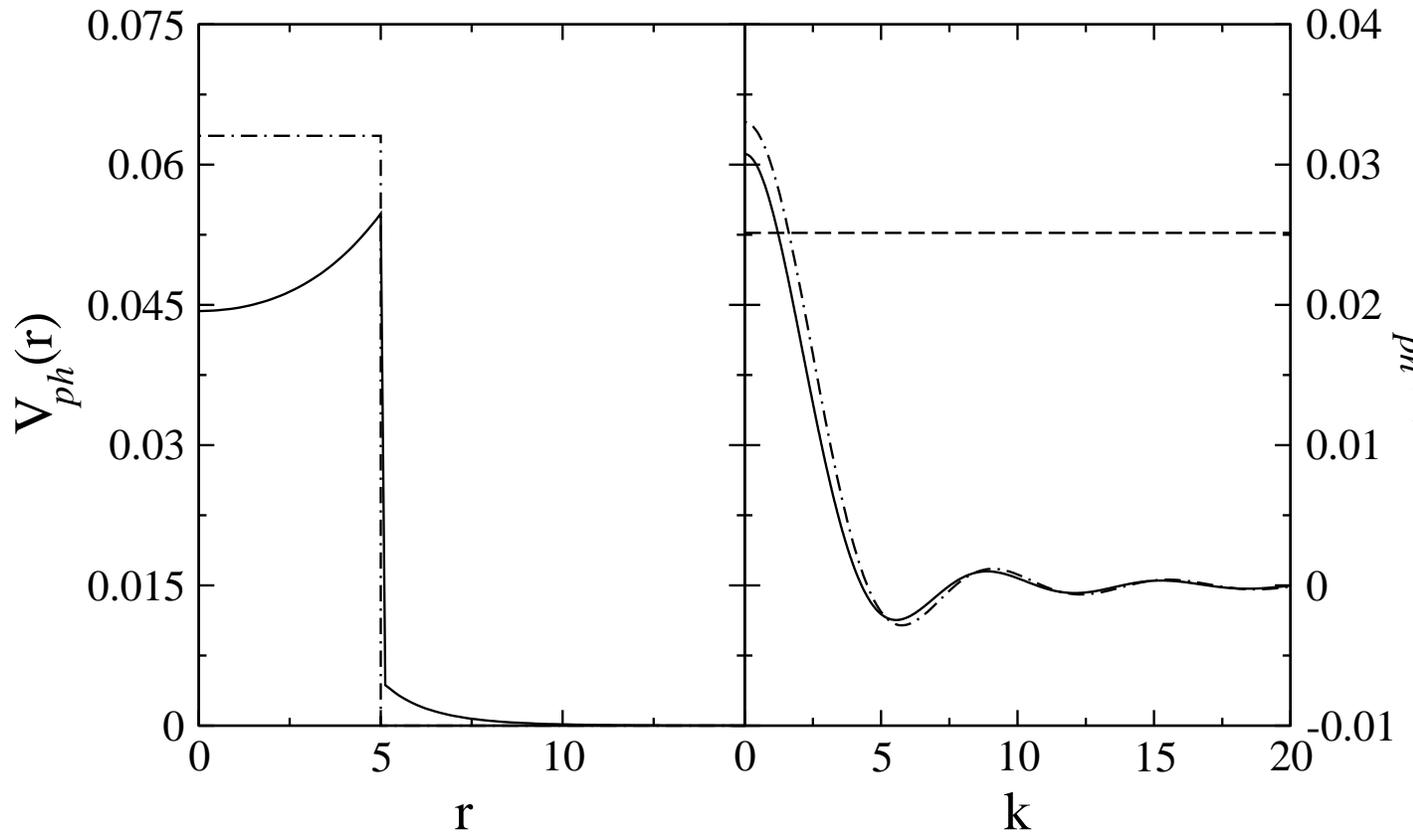}
\end{center}
\caption{Particle-hole interaction in r--space (left panel) and in 
k--space (right panel), for SS5 at $x=0.001$. 
The solid and dot--dashed lines correspond to the El and UL 
cases, respectively. 
The dashed line in the right panel gives the Bogoliubov $V_{ph}$. 
Distances in units of $a$ and momenta in units of $a^{-1}$.  
}
\label{fig-vphssx0}
\end{figure}

The particle-hole interaction corresponding to SS5 at
$x=0.001$ is shown in Fig.~\ref{fig-vphssx0} 
for the EL, UL and Bogoliubov approaches.  The left and right 
panels give $V_{ph}(r)$ and its Fourier transform, $V_{ph}(k)$, respectively. 
$V_{ph}^{B}(r)=8 \pi x \delta({\bf r})$ is not
shown in the figure. The {\em ph}-interaction in the UL limit
coincides with the interaction itself, 
$V_{ph}^{UL}(r)=V(r)$.  $V_{ph}^{EL}(r)$ 
 is discontinuous at $r=R$. The differences
between $V_{ph}^{UL}(k)$ and $V_{ph}^{EL}(k)$ are larger 
at low momenta, consistently with the differences observed in
the static structure function. 

\begin{figure}
\begin{center}
\includegraphics*[height=11cm]{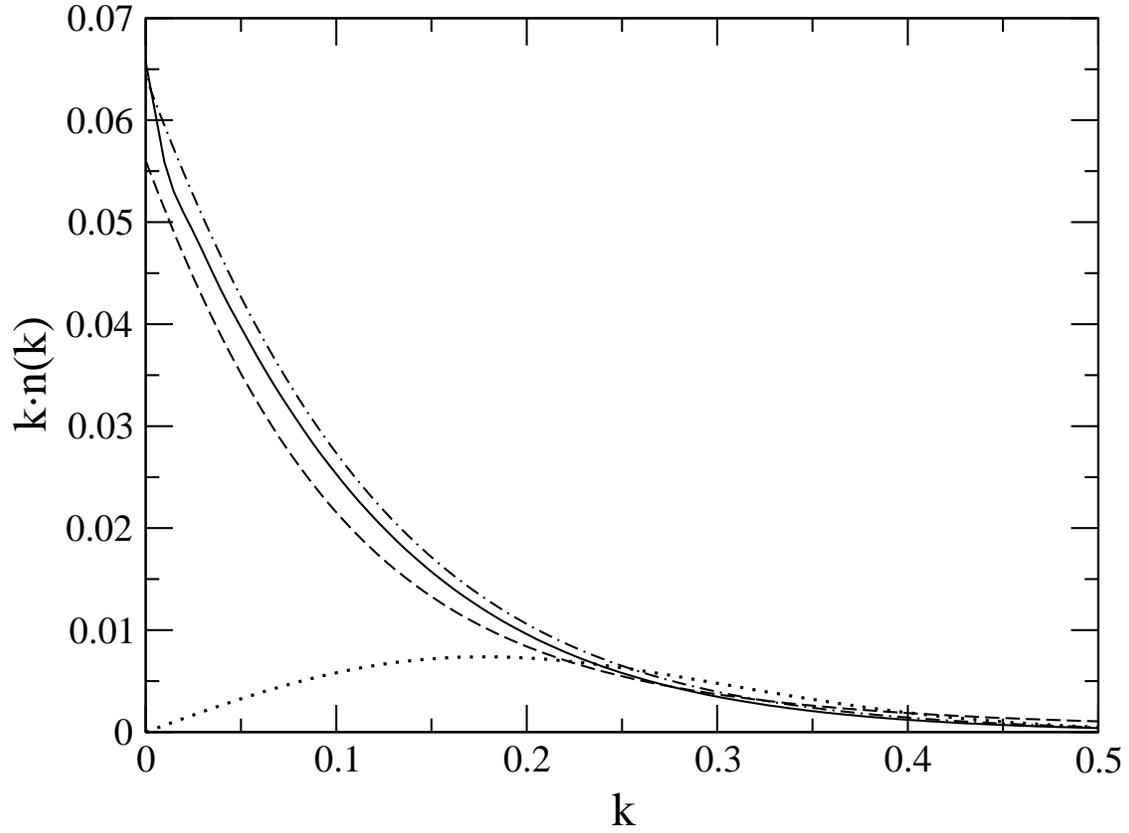}
\end{center}
\caption{$k\,n(k)$ for SS5 at $x=0.001$. 
The solid, dashed, dash--dotted and dotted lines correpond to 
the EL, Bogoliubov, UL and SR results respectively. 
Momenta in units of $a^{-1}$.  
}
\label{fig-knkall}
\end{figure}

 Finally  we discuss the results for the momentum distribution, $n(k)$. 
The quantity $k\,n(k)$ at $x=0.001$ is plotted in Fig.~\ref{fig-knkall} 
for the SS5 interaction in the EL, Bogoliubov, UL and SR cases. 
All  momentum distributions  but the short-range  one  satisfy the long--wavelength behavior 
(\ref{knk0EL}). 
 Since $n_{SR}(k=0)\neq 0$, $k\,n_{SR}(k)=0$ at the origin. 
As in the HS case,  $k\,n_{SR}(k)$ presents a
maximum at intermediate momenta, as a byproduct of the 
absence of a long--range structure in the two--body correlation factor
$f_{SR}(r)$. 

In the $x$-range  considered,  $k\,n_{EL}(k)$ is a monotonously
decreasing function of $k$. At higher $x$ it develops a maximum at 
low momenta, consistently with the HS results,  
 also  found in the high density homogenous 
atomic $^4$He. In the UL, 
\begin{equation}
\lim_{k \rightarrow 0} {k\,n_{UL}(k)}  = 
\frac {c_{UL}}{4} - \frac {k}{2} \, ,
\label{nkasymp}
\end{equation}
and the value at the origin depends on $x$ through the sound velocity, 
whereas its slope, $-1/2$, is independent on the density.

$k\,n_{B}(k)$ is also  a decreasing function of $k$, its $k=0$ 
value being slightly smaller than the UL one, corresponding to 
a lower sound velocity. The UL momentum
distribution for the SS potential decays as $k^{-8}$, and the condensate
fraction can be computed by imposing the normalization condition.
The condensate fractions and the kinetic energies obtained from 
the momentum distribution are reported in Table~\ref{tab-3}.

\begin{figure}
\begin{center}
\includegraphics*[height=11cm]{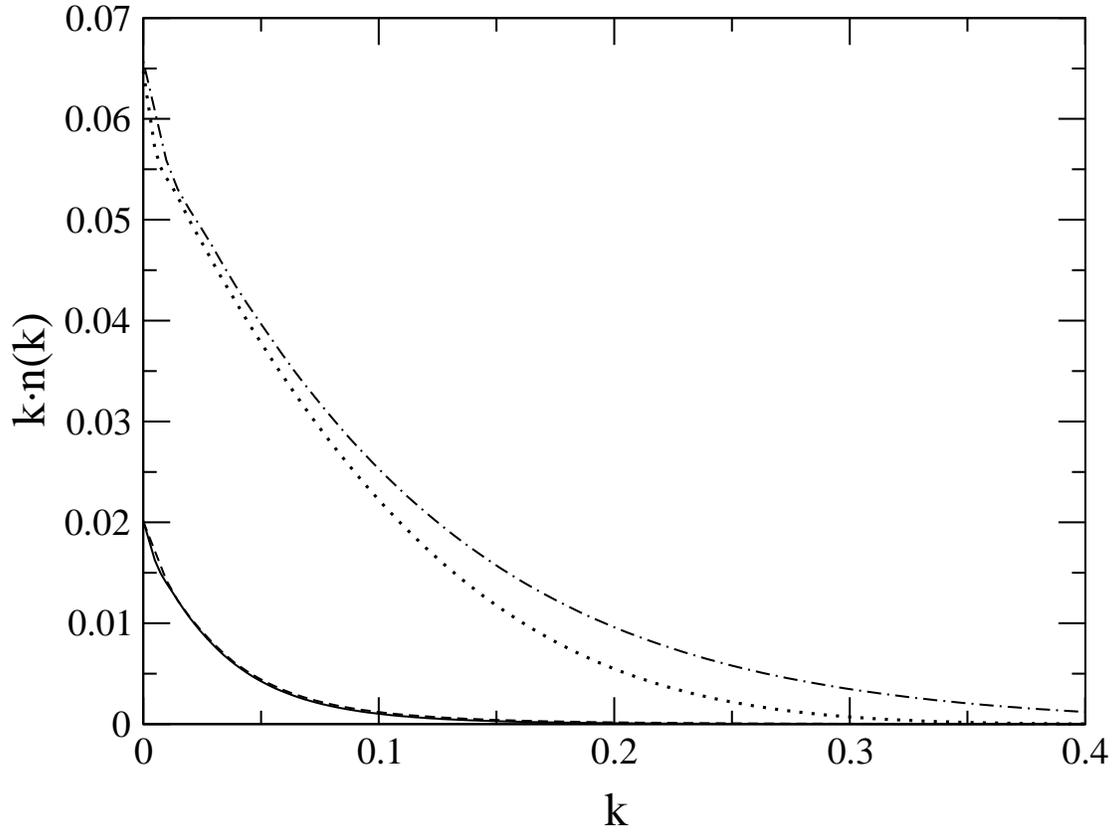}
\end{center}
\caption{EL $k\,n(k)$ at 
$x=0.0001$ (solid line for SS10, dashed line for SS5) and 
$x=0.001$ (dotted line for SS10, dot--dashed line for SS5). 
Momenta in units of $a^{-1}$.  
}
\label{fig-knkSSEL}
\end{figure}

The dependence of $n(k)$ on the shape of the potential is studied in 
Fig.~\ref{fig-knkSSEL}, where $k\,n_{EL}(k)$ is plotted at
$x=0.0001$ and $x=0.001$ for SS5 and SS10.
 Apparently, the SS5 and the SS10 momentum distributions are identical at  
$x=0.0001$, pointing to a dependence only on the scattering
length. This conclusion must be taken with caution, since in 
Table~\ref{tab-3} more than a factor of two between the SS5 
and SS10 kinetic energies, ${\bar T}_n$ , is found even at this 
low density. 

\begin{figure}
\begin{center}
\includegraphics*[height=11cm]{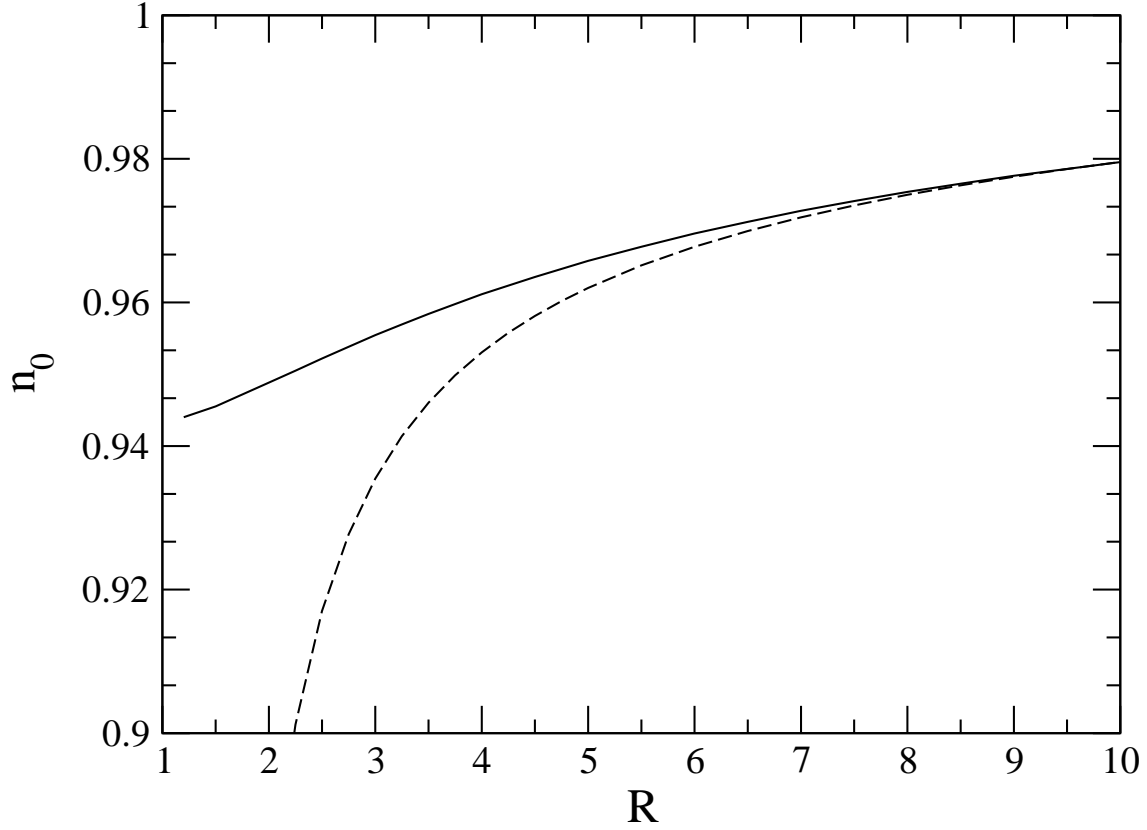}
\end{center}
\caption{Condensate fraction for the soft sphere gas, 
at $x=0.001$, as a function of the radius, $R$, of the 
SS potential at fixed scattering length.  
EL, solid line; UL, dashed line. The Bogoliubov condensate 
fraction is $n_{0B}=0.952$.  Distances in units of $a$.  
} 
\label{fig-n0ssx0.001}
\end{figure}

The condensate fraction also depends on the shape of the potential. 
Figure~\ref{fig-n0ssx0.001} shows $n_{0,EL}$ and
$n_{0,UL}$ at $x=0.001$ as a function of the radius of the SS 
potential, $R$, at a fixed value of the scattering length.  As expected,
the condensate fraction grows with $R$, since the interaction 
softens. The UL approach becomes accurate at large $R$--values, in the 
very weak interaction limit.
In all cases, $n_{0,UL}$ is smaller than $n_{0,EL}$. 

\section{Summary and Conclusions}

 We have carefully investigated the energy and structure of a 
homogeneous gas of bosons interacting via hard and soft 
sphere potentials. 
We have adopted and compared several many-body approaches, as 
the variational correlated theory, the Bogoliubov model and the 
 uniform limit approximation, valid in the weak interaction regime. 
When possible, the results have been compared with 
the exact diffusion Monte Carlo ones. A Jastrow type correlation  
appears to produce a good quality wave function 
 if the hypernetted chain energy functional is freely 
minimized and the resulting Euler equation is solved. 
This is true for both hard and soft sphere interactions. 
The study of soft sphere potentials has confirmed the appearence of a shape 
dependence in the energy at $x\sim 0.001$, as already found by 
the IPC calculations of Ref.\cite{sarsa02} and by the DMC 
study of Ref.\cite{boro99}. 
We have numerically compared the EL results with those obtained 
in the uniform limit of weak interaction, 
in the Bogoliubov approximation and  in the IPC theory. 
As expected, the differences are more relevant 
for the strongest interactions. The uniform-limit and independent-pair-correlation
 energies become 
more and more reliable as the interaction weakens. 
The universality breaks at much lower $x$--values for quantities 
different from the energy. For instance, the short range structure 
of the radial distribution functions (and, in consequence, the 
large momentum behavior of the static structure function) largely 
depends on the shape of the potential already at $x=0.0001$. 
We find a potential shape dependence in the SS condensate fraction at 
$x=0.001$. These results, as well as those for the energy, may help 
in evaluating the parameters entering the nonuniversal corrections 
in effective field theory\cite{braaten01}, whose extraction from the 
available DMC calculations is heavily biased by the statistical 
errors.

The presence of a maximum in the correlated distribution function when 
$x$ increases has been discussed in details. The Bogoliubov model 
does not show this short range structure, independently on $x$. 
The excitation spectrum develops a rotonic maximum at large $x$, 
related to the shape of the particle--hole interaction. Again, 
the Bogoliubov approach, providing a constant $V_{ph}(k)$, does not 
allow for a roton--like excitation. The effect of the correlations 
is also apparent in the momentum distributions. In the hard sphere gas 
the CBF $kn(k)$ acquires a maximum at low momentum at $x\sim 0.06$. 
Both the Bogoliubov and the uniform limit (for soft spheres) approaches 
fail to reproduce this behavior. 

The correlated basis functions theory may be extended to treat fermionic 
hard and soft spheres or mixtures of Fermi--Bose gases. Work along these 
lines is in progress. More challenging is the application of the 
CBF approach to trapped atomic gases, without resorting to 
local density type approximations\cite{fabro99}. 
Employing the variational method in finite dilute 
systems presents mainly technical difficulties. However, it 
will probably represent the natural development of this 
technique.

\acknowledgments

The authors are grateful to Profs. S.Fantoni and J.Boronat 
for many useful discussions and to Prof. S.Giorgini and Dr.
A.Sarsa for providing their DMC and IPC results, respectively. 
This work has been partially supported by Grant 
No. BFM2002-01868 from DGI (Spain),  
Grant No. 2001SGR-00064 from  the Generalitat de Catalunya,  
and by the Italian MIUR through the {\it PRIN: Fisica Teorica 
del Nucleo Atomico e dei Sistemi a Molti Corpi}.

\pagebreak

\pagebreak

\begin{table}[!t]
\begin{center}
{
\begin{tabular}{cccccc}
$x$& ${\bar E}_{DMC}/N$ & ${\bar E}_{EL}/N$ & ${\bar E}_{SR}/N$ & $d$ & 
${\bar E}_{LD}/N$\\ \hline
$10^{-6}$       & $1.262\cdot 10^{-5}$ & $1.264\cdot 10^{-5}$ &
$1.261\cdot 10^{-5}$ & $480$ & $1.260\cdot10^{-6}$ \\ 
$10^{-5}$       & $1.274\cdot 10^{-4}$ & $1.279\cdot 10^{-4}$ &
$1.277\cdot 10^{-4}$ & $140$ & $1.276\cdot10^{-4}$ \\ 
$10^{-4}$       & $1.311\cdot 10^{-3}$ & $1.316\cdot 10^{-3}$ &
$1.313\cdot 10^{-3}$ & $90$ & $1.317\cdot10^{-3}$ \\ 
$10^{-3}$       & $1.424\cdot 10^{-2}$ & $1.430\cdot 10^{-2}$ &
$1.428\cdot 10^{-2}$ & $34$ & $1.448\cdot10^{-2}$ \\ 
$5\cdot10^{-3}$ & $8.155\cdot 10^{-2}$ & $8.206\cdot 10^{-2}$ &
$8.217\cdot 10^{-2}$ & $15$ & $8.422\cdot10^{-2}$ \\ 
$10^{-2}$       & $1.796\cdot 10^{-1}$ & $1.814\cdot 10^{-1}$ &
$1.819\cdot 10^{-1}$& $11$ & $1.862\cdot10^{-1}$ \\ 
$5\cdot10^{-2}$ & $1.338$              & $1.383$		&
$1.402$	& $7$ & $1.305$ \\
$0.1$		& $3.627$		& $3.848$		&
$3.971$	& $6$ & $3.170$ \\ 
\end{tabular}
}
\end{center}
\caption{Scaled energy per particle for the hard spheres model, 
 as a function of $x$. See text. $d$ is the healing distance 
corresponding to the correlation factor of Eq.(\ref{fr-b1}), in 
units of the scattering length. In ${\bar E}_{LD}$ the first two 
terms of the expansion (\ref{E_LY}) have been used.}
\label{tab-1}
\end{table}

\pagebreak

\begin{table}[!h]
\begin{center}
{
\begin{tabular}{ccccccc}
&$x$   &${\bar T}_g/N$&${\bar T}_n/N$&$n_0$&$n_{0DMC}$&$n_{0B}$\\ \hline
EL  & 0.001 & $1.430\cdot 10^{-2}$ & $1.270\cdot 10^{-2}$ & 0.947&0.948&0.952\\
SR  & 0.001 & $1.428\cdot 10^{-2}$ & $1.290\cdot 10^{-2}$ & 0.940 &  &  \\
EL  & 0.01  & $1.814\cdot 10^{-1}$ & $1.772\cdot 10^{-1}$ & 0.801 &0.803&0.850\\
SR  & 0.01  & $1.819\cdot 10^{-1}$ & $1.975\cdot 10^{-1}$ & 0.799 &  &  \\
EL  & 0.05  & 1.383  & 1.594 & 0.493 & 0.501 & 0.664 \\
SR  & 0.05  & 1.402  & 1.543 & 0.481 &       &       \\
EL  & 0.08  & 2.728  & 3.414 & 0.340 &  & 0.574 \\
SR  & 0.08  & 2.794  & 3.451 & 0.327 &       &       \\
\end{tabular}
}
\end{center}
\caption{Scaled kinetic energy per particle of the HS gas calculated 
using the momentum distribution and the radial distribution function 
for the EL and SR wave functions at several values of $x$. 
$n_0$, $n_{0DMC}$ and $n_{0B}$ are the EL, DMC and Bogoliubov condensate 
fractions.}
\label{tab-2}
\end{table}

\pagebreak

\begin{table}[!h]
\begin{center}
{
\begin{tabular}{clllllll}
&$x$&$R$&${\bar E}/N$ &${\bar V}/N$ &${\bar T}_g/N$&${\bar T}_n/N$ & $n_0$
\\ \hline \hline
EL  & $10^{-4}$ & 10 &$1.305\cdot 10^{-3}$&$1.202\cdot 10^{-3}$& 
                      $1.038\cdot 10^{-4}$&$1.000\cdot 10^{-4}$ & 0.988 \\
SR  & $10^{-4}$ & 10 &$1.317\cdot 10^{-3}$&$1.241\cdot 10^{-3}$& 
                      $0.765\cdot 10^{-4}$&$0.765\cdot 10^{-4}$ & 0.997 \\
UL  & $10^{-4}$ & 10 &$1.295\cdot 10^{-3}$&$1.184\cdot 10^{-4}$& 
                      &$1.110\cdot 10^{-4}$ & 0.992 \\
IPC & $10^{-4}$ & 10 &$1.311\cdot 10^{-3}$& & & &  \\ 
DMC & $10^{-4}$ & 10 &$1.303\cdot 10^{-3}$& & & & 0.989 \\ \hline
EL  & $10^{-4}$ & 5  &$1.314\cdot 10^{-3}$&$1.044\cdot 10^{-3}$& 
                      $2.708\cdot 10^{-4}$&$2.630\cdot 10^{-4}$ & 0.985 \\
SR  & $10^{-4}$ & 5  &$1.361\cdot 10^{-3}$&$1.138\cdot 10^{-3}$& 
                      $2.231\cdot 10^{-4}$&$2.231\cdot 10^{-4}$ & 0.996 \\
UL  & $10^{-4}$ & 5  &$1.231\cdot 10^{-3}$&$0.853\cdot 10^{-3}$& 
                      &$3.780\cdot 10^{-4}$ & 0.982 \\
IPC & $10^{-4}$ & 5  &$1.331\cdot 10^{-3}$& & & & 0.987 \\ 
DMC & $10^{-4}$ & 5  &$1.309\cdot 10^{-3}$& & & &0.989  \\ \hline
EL  & $10^{-2}$ & 10 &$1.404\cdot 10^{-1}$&$1.394\cdot 10^{-1}$& 
                      $0.990\cdot 10^{-3}$&$0.981\cdot 10^{-3}$ & 0.980 \\
SR  & $10^{-2}$ & 10 &$1.405\cdot 10^{-1}$&$1.395\cdot 10^{-1}$& 
                      $0.960\cdot 10^{-3}$&$0.960\cdot 10^{-3}$ & 0.977 \\
UL  & $10^{-2}$ & 10 &$1.404\cdot 10^{-1}$&$1.395\cdot 10^{-1}$& 
                      &$0.963\cdot 10^{-3}$ & 0.980 \\
IPC & $10^{-2}$ & 10 &$1.408\cdot 10^{-1}$& & & &  \\ \hline
EL  & $10^{-2}$ & 5  &$1.532\cdot 10^{-1}$&$1.468\cdot 10^{-1}$& 
                      $6.445\cdot 10^{-3}$&$6.480\cdot 10^{-3}$ & 0.951 \\
SR  & $10^{-2}$ & 5  &$1.535\cdot 10^{-1}$&$1.481\cdot 10^{-1}$& 
                      $5.350\cdot 10^{-3}$&$5.350\cdot 10^{-3}$ & 0.960 \\
UL  & $10^{-2}$ & 5  &$1.528\cdot 10^{-1}$&$1.464\cdot 10^{-1}$& 
                      &$6.375\cdot 10^{-3}$ & 0.950 \\
IPC & $10^{-2}$ & 5  &$1.556\cdot 10^{-1}$& & & & 0.953 \\ 
\end{tabular}
}
\end{center}
\caption{Scaled energies and condensate fractions  for the soft sphere gas 
 in the El, SR, UL, IPC and DMC approaches for two potentials having the same 
 scattering length (see text) at $x=10^{-4}$ and $x=10^{-2}$.}
\label{tab-3}
\end{table}

\pagebreak


\pagebreak

\pagebreak

\pagebreak

\pagebreak

\pagebreak

\pagebreak

\pagebreak
\pagebreak

\pagebreak

\pagebreak

\pagebreak

\pagebreak

\end{document}